\documentclass[11pt]{article}
\usepackage{latexsym}
\usepackage{amssymb}
\usepackage{graphics}
\usepackage{epsfig}
\usepackage{rotating}
\usepackage{amsfonts}
\usepackage{amsmath}
\usepackage{hyperref}
\hypersetup{
	colorlinks=true,
	linkcolor=blue,
	filecolor=magenta,
	urlcolor=cyan,
}
\begin{document}

\begin{titlepage}
\begin{flushright}
IRMP-CP3-23-07\\
\end{flushright}

\vspace{5pt}

\begin{center}

{\Large\bf Magnetic Monopoles with No Strings Attached:}\\

\vspace{7pt}

{\Large\bf A Portal to the Dark Side of Dual Electrodynamics}\\

\vspace{80pt}

Jan Govaerts$^{a,b,}$\footnote{Fellow of the Stellenbosch Institute for Advanced Study (STIAS), Stellenbosch,
Republic of South Africa}

\vspace{30pt}

$^{a}${\sl Centre for Cosmology, Particle Physics and Phenomenology (CP3),\\
Institut de Recherche en Math\'ematique et Physique (IRMP),\\
Universit\'e catholique de Louvain (UCLouvain),\\
2, Chemin du Cyclotron, B-1348 Louvain-la-Neuve, Belgium}\\
E-mail: {\em Jan.Govaerts@uclouvain.be}\\
ORCID: {\tt http://orcid.org/0000-0002-8430-5180}\\
\vspace{15pt}
$^{b}${\sl International Chair in Mathematical Physics and Applications (ICMPA--UNESCO Chair)\\
University of Abomey-Calavi, 072 B.P. 50, Cotonou, Republic of Benin}\\

\vspace{10pt}


\vspace{20pt}

\begin{abstract}
\noindent
It has long been known that in the absence of electric charges and currents, Maxwell's electromagnetism
in 4 dimensional vacuum Minkowski space-time is invariant under SO(2) dual transformations that mix its electric and magnetic fields.
Extending this symmetry to include the coupling to electrically charged matter, requires a dual coupling to magnetically charged matter as well,
leading to Maxwell equations for SO(2) dual electrodynamics. Based on a doubled ensemble of SO(2) dual 4-vector gauge potentials which does
away with the need of Dirac string singularities for magnetic monopoles, a local Lagrangian action principle for SO(2) dual electromagnetism is known,
which manifestly displays all the required space-time and internal symmetries, and reduces to the experimentally well established Maxwell electrodynamics
in the absence of magnetic charges and currents. Applying the same considerations for the matter action of electrically and magnetically charged
point particles, a unique SO(2) dual generalised Lorentz force is identified for SO(2) dual electrodynamics, truly different from the usual SO(2) dual invariant
choice motivated by simplicity, but yet made arbitrarily and which does not derive from some action principle.
This generalised Lorentz force involves a single real and new coupling constant of unknown value, without the requirement of a Dirac-Schwinger-Zwanziger
quantisation condition for electric and magnetic charges of dyons. A physical consequence for SO(2) dual electrodynamics of this coupling constant
\textcolor{black}{if nonvanishing},
is to open a channel, or portal between the otherwise mutually totally ``dark'' sectors of electric and magnetic charges for electromagnetic interactions.
\end{abstract}

\end{center}

\end{titlepage}

\setcounter{footnote}{0}

\section{Introduction}
\label{Intro}

In 4 dimensional Minkowski space-time, consider the ordinary vacuum Maxwell equations with electric sources,
\begin{equation}
\vec{\nabla}\cdot\frac{\vec{E}}{c}=\mu_0(c\rho_e),\qquad
\vec{\nabla}\times\frac{\vec{E}}{c}+\frac{\partial}{\partial(ct)}\vec{B}=\vec{0},\qquad
\vec{\nabla}\cdot\vec{B}=0,\qquad
\vec{\nabla}\times\vec{B}-\frac{\partial}{\partial(ct)}\frac{\vec{E}}{c}=\mu_0\,\vec{J}_e,
\label{eq:Maxwell}
\end{equation}
as well as the relativistic Lorentz force equation of motion for a massive point charge of mass $m$ and electric charge $q$,
\begin{equation}
\frac{d}{dt}\vec{p}=qc\left(\frac{\vec{E}}{c}+\frac{\vec{v}}{c}\times\vec{B}\right),\qquad
\vec{p}=\frac{m\vec{v}}{\sqrt{1-\vec{v}\,^2/c^2}},\qquad
\vec{v}=\frac{d\vec{x}}{dt},
\label{eq:Lorentz}
\end{equation}
in manifestly 3 dimensional Euclidean covariant notations that certainly are standard and this for physical quantities expressed in S.I. units,
with in particular $\epsilon_0 \mu_0 c^2=1$ \cite{Jackson}. The electric charge and current densities as sources for the Maxwell equations,
$\rho_e$ and $\vec{J}_e$, respectively, then necessarily obey the continuity equation for the conservation of the total electric charge,
\begin{equation}
\frac{\partial }{\partial t}\rho_e + \vec{\nabla}\cdot\vec{J}_e=0,\qquad
\frac{d}{dt}Q_e=0,\qquad Q_e=\int_{(\infty)} d^3\vec{x}\,\rho_e(ct,\vec{x}\,).
\end{equation}

The ensemble of this set of coupled equations is solidly grounded in many well established experimental observations dating back to the XVIII$^{\rm th}$
and XIX$^{\rm th}$ centuries (see for instance Refs.\cite{Jackson,Babel}).
Once Maxwell had included the displacement current term $(\mu_0\epsilon_0\partial\vec{E}/\partial t)$
into Amp\`ere's Law to comply with electric charge conservation, this most beautiful physics conceptual edifice was made complete, with the theoretical prediction
for the existence of electromagnetic waves propagating with the velocity $1/\sqrt{\epsilon_0\mu_0}=c$, whose reality was to be confirmed experimentally
sometime later.

Among the many lessons these equations taught and are still teaching us, one must count the discovery of (special) Lorentz transformations,
hence the conceptual paradigm of Einstein's Special Relativity and Minkowski space-time, of which the physics relevance need not be emphasized.
The relative negative
sign in the signature for the Minkowski space-time metric lies hidden in these equations, as is revealed for instance in the wave equations implied
for the electric and magnetic fields from Maxwell's equations,
\begin{equation}
\left(\frac{\partial^2}{\partial(ct)^2}\,-\,\vec{\nabla}^2\right) \frac{\vec{E}}{c}=-\mu_0\vec{\nabla}(c\rho_e)-\mu_0\frac{\partial}{\partial (ct)}\vec{J}_e,\qquad
\left(\frac{\partial^2}{\partial(ct)^2}\,-\,\vec{\nabla}^2\right) \vec{B}=\mu_0\vec{\nabla}\times\vec{J}_e.
\end{equation}

However, an understanding of the actual meaning
of yet another global symmetry underlying Maxwell's equations has remained wanting ever since when, for more than a century now,
Heaviside \cite{Jose} pointed to a dual symmetry between the electric and magnetic sectors as represented by these equations,
at least in the absence of sources. Indeed when both $\rho_e=0$ and $\vec{J}_e=\vec{0}$, the sourceless Maxwell equations
are covariant under the discrete duality exchange transformation,
\begin{equation}
\frac{\vec{E}}{c}\longrightarrow \frac{\vec{E}'}{c}=\vec{B},\qquad
\vec{B} \longrightarrow \vec{B}'=-\frac{\vec{E}}{c}.
\end{equation}
More generally the sourceless Maxwell equations are covariant under the following continuous global SO(2) duality transformations for whatever value of the
constant angular variable $\theta$, with the value $\theta=\pi/2$ corresponding to the above discrete duality transformation,
\begin{equation}
\left(\begin{array}{c}
\vec{E}'/c \\
\vec{B}'
\end{array}\right)=
\left(\begin{array}{c c}
\cos\theta & \sin\theta \\
-\sin\theta & \cos\theta
\end{array}\right)
\cdot\left(\begin{array}{c}
\vec{E}/c \\
\vec{B}
\end{array}\right)=
\left(\begin{array}{c}
\cos\theta\,\vec{E}/c\,+\,\sin\theta\,\vec{B} \\
-\sin\theta\,\vec{E}/c\,+\,\cos\theta\,\vec{B}
\end{array}\right).
\label{eq:SO2-EB}
\end{equation}

By hiding in Maxwell's equations another such continuous symmetry, is Dame Nature wanting to reveal yet another profound truth about
the physical world not directly accessible to neither our senses nor our detectors?
Whatever the case may be, over all these decades this electromagnetic duality symmetry has never ceased
to remain not only a topic of fascinating interest, but has provided as well nonperturbative means for important insights into the dynamics and spectral content of
generalised nonabelian Yang-Mills gauge theories with their spontaneous symmetry breaking mechanisms and thereby \textcolor{black}{topological} monopole solutions,
as well as their extensions to string and $M$-theories (see for instance Refs.\cite{Jose,Alvarez,Shnir}).

Taking in earnest such a clue from fundamental physics laws and thus wanting to extend the SO(2) electromagnetic duality to situations when
electric sources $\rho_e$ and $\vec{J}_e$ are active as well, necessarily requires a dual charge sector, namely
magnetic charge and current densities, $\rho_m$ and $\vec{J}_m$, respectively, which obey as well a continuity equation for total magnetic charge conservation,
\begin{equation}
\partial_t\rho_m+\vec{\nabla}\cdot\vec{J}_m=0,\qquad
\frac{d}{dt}Q_m=0,\qquad Q_m=\int_{(\infty)} d^3\vec{x}\,\rho_m(ct,\vec{x}\,),
\end{equation}
but which would have heretofore remained undetected because of a lack of (sufficiently strong)
interactions with purely electrically charged matter of densities $\rho_e$ and $\vec{J}_e$.

Furthermore, SO(2) duality transformations are then extended to act as follows on the electric and magnetic sources,
\begin{equation}
\left(\begin{array}{c}
c\rho_e' \\
c\rho_m'
\end{array}\right)=
\left(\begin{array}{c c}
\cos\theta & \sin\theta \\
-\sin\theta & \cos\theta
\end{array}\right)\cdot
\left(\begin{array}{c}
c\rho_e \\
c\rho_m
\end{array}\right),\quad
\left(\begin{array}{c}
\vec{J}_e^{\,'} \\
\vec{J}_m^{\,'}
\end{array}\right)=
\left(\begin{array}{c c}
\cos\theta & \sin\theta \\
-\sin\theta & \cos\theta
\end{array}\right)\cdot
\left(\begin{array}{c}
\vec{J}_e \\
\vec{J}_m
\end{array}\right),
\end{equation}
while the extended Maxwell equations of SO(2) dual electromagnetism then read,
\begin{equation}
\vec{\nabla}\cdot\frac{\vec{E}}{c}=\mu_0(c\rho_e),\quad
\vec{\nabla}\times\frac{\vec{E}}{c}+\frac{\partial}{\partial(ct)}\vec{B}=-\mu_0\vec{J}_m,\quad
\vec{\nabla}\cdot\vec{B}=\mu_0(c\rho_m),\quad
\vec{\nabla}\times\vec{B}-\frac{\partial}{\partial(ct)}\frac{\vec{E}}{c}=\mu_0\,\vec{J}_e,
\label{eq:DualEM}
\end{equation}
which are indeed manifestly SO(2) dual covariant.

However, what is then to be done of the above Lorentz force equation of motion (\ref{eq:Lorentz})? 
The most direct though naive SO(2) dual invariant choice would be to extend it
in the following way, now for a massive point charge of mass $m$, electric charge $q$ and magnetic charge $g$\footnote{Note that given the present choice of units,
each of $\rho_m$, $\vec{J}_m$ and $g$ has the same physical dimension as each of $\rho_e$, $\vec{J}_e$ and $q$, respectively.},
\begin{equation}
\frac{d}{dt}\vec{p}=qc\left(\frac{\vec{E}}{c}+\frac{\vec{v}}{c}\times\vec{B}\right)\,+\,
gc\left(\vec{B}-\frac{\vec{v}}{c}\times\frac{\vec{E}}{c}\right)\,+\,? ,
\label{eq:generLorentz}
\end{equation}
where nonetheless the question mark indicates that possibly, further SO(2) dual invariant terms may be required following a more thorough analysis.
Yet the choice that seems to be always made in the (very vast) literature (on the subject) without questioning it, is the minimal one of only considering
the sum of the four terms given explicitly on the r.h.s.~of this latter expression.

Indeed, even though the above equations in (\ref{eq:DualEM}) and (\ref{eq:generLorentz}) are manifestly SO(2) dual covariant
for the SO(2) transformations defined above\footnote{And of course by including as well the transformations for the electric and magnetic point charges, namely
$q'=q\cos\theta\,+\,g\sin\theta$ and $g'=-q\sin\theta\,+\,g\cos\theta$.},
none of the additional contributions in these generalised equations have ever been subjected to any experimental pronouncement by Dame Nature,
for the simple reason that magnetic monopole charges have never been observed, certainly not through (even quantum mechanical)
electromagnetic interactions with ordinary (and thus only electrically charged) matter.
The choice of Maxwell equations for dual electromagnetism in (\ref{eq:DualEM}) is certainly the minimal and simplest one possible.
As is always done in the literature, it will be taken for granted here as well in the form (\ref{eq:DualEM}) given above---it certainly is difficult to think of any other.
The issue of the Lorentz force however, is a different matter.

For one thing, \textcolor{black}{it is known \cite{Rohr}}
that such an equation of motion in exactly the form of (\ref{eq:generLorentz}) (without any extra terms) cannot derive
from an action principle\footnote{This fact is established \textcolor{black}{again} in this work.},
which makes at least for an uneasy situation when it comes to a fundamental interaction.
Furthermore, it should be noted that since SO(2) possesses two invariant tensors of rank 2, namely
$\delta_{ab}=\delta^{ab}$ and $\epsilon_{ab}=\epsilon^{ab}=-\epsilon_{ba}$ with $a,b=e,m$ and $\epsilon_{em}=+1$,
given any two SO(2) vectors, say $\phi^a_1$ and $\phi^a_2$,
they define two independent SO(2) invariants, namely $\delta_{ab}\phi^a_1\phi^b_2$ and $\epsilon_{ab}\phi^a_1\phi^b_2$. Consequently there may
exist still other SO(2) dual invariant terms that could come to replace the question mark in (\ref{eq:generLorentz}), and then produce a Lorentz force
equation of motion that could even derive from some action principle.

On the other hand, bringing into the discussion the possibility of magnetic monopole charges also raises the issue of the Dirac strings of magnetic flux
these monopoles would carry along \cite{Dirac1}, at least in an approach that would insist on representing the electric and magnetic fields in terms
of the usual single pair of scalar and vector potentials, $\Phi$ and $\vec{A}$, such that
\begin{equation}
\vec{E}=-\vec{\nabla}\Phi - \partial_t\vec{A},\qquad
\vec{B}=\vec{\nabla}\times\vec{A},
\end{equation}
thus even when $\vec{\nabla}\cdot\vec{B}=\mu_0(c\rho_m)\ne 0$
(while mathematics certainly requires that $\vec{\nabla}\times(\vec{\nabla}\times\vec{A})=\vec{0}$ for any analytic vector field $\vec{A}$\,).
As Dirac has famously shown \cite{Dirac1,Dirac2}, any such vector potential must possess a nonlocal singularity
structure along (at least) one string connecting the magnetic monopole to the point at infinity. However, at the quantum level this singularity may be
made invisible to some particle of electric charge $q$ provided the product of that electric charge with the monopole's magnetic charge obeys
a specific quantisation rule proportional to Planck's reduced constant $\hbar$ \cite{Dirac1,Dirac2,Schwinger,Zwanziger}.
Nevertheless the presence of such nonlocal time-like string singularities in space
makes again at least for an uneasy situation\footnote{Notwithstanding the fact that a combination of some multi-covering of 3-Euclidean space and
appropriately correlated gauge transformations may evade this unwelcome feature \cite{Wu,Jose}.}, in particular when considering the action principle for the dynamics
of the gauge potentials $\Phi$ and $\vec{A}$.
And furthermore, an action which itself, when expressed within the Lagrangian formalism at least, cannot be made manifestly SO(2) dual invariant,
in contradistinction to its space-time Poincar\'e and Lorentz symmetries.

The present work addresses these different issues, by taking at face value the following list of perfectly reasonable and acceptable
criteria or requirements, as well as their consequences, for what is indeed a fundamental interaction in its classical realm.
The dynamics of electromagnetic phenomena coupling to electrically
and magnetically charged matter, as well as the dynamics of the latter, should all derive from some action principle. This action principle should be local
in space-time, without any field displaying some nonlocal singularity of a Dirac string type. This action even in Lagrangian form should be manifestly invariant
not only under Poincar\'e and Lorentz transformations, but under the above SO(2) electromagnetic duality transformations as well. Finally besides
these requirements of a theoretical character, in order to stray away as little as possible from experimental orthodoxy and its solidly established facts,
the construction should be
such that when the magnetically charged matter is absent, namely when $\rho_m=0$ and $\vec{J}_m=\vec{0}$, and when the test charges carry
only an electric charge, $q\ne 0$, but no magnetic charge whatsoever, $g=0$, the electromagnetic and Lorentz force equations of motion (\ref{eq:DualEM})
and (\ref{eq:generLorentz}) must reduce back to the ordinary expressions in (\ref{eq:Maxwell}) and (\ref{eq:Lorentz}) which have successfully met
Dame Nature's verdict, since indeed all experimental electromagnetic observations have until now been effected only under such circumstances.

\textcolor{black}{
The present discussion adds one more work with its own specific approach,
 to a large body indeed of literature dealing with all those fascinating issues prompted
by the possible existence of magnetic monopoles (see for instance the reviews in Refs.\cite{Blago,Singleton,Milton}). However it differs from such
published work in two main respects. On the one hand, its insistence of having dynamical equations of motion that all derive from an action
principle which even already at the classical level, is manifestly Poincar\'e, Lorentz, U(1) gauge and SO(2) dual invariant, as well as local in space-time.
And on the other hand, such that the generalised SO(2) duality covariant Lorentz force equation for electric and magnetic point charges subjected to such SO(2) dual
covariant electrodynamics not be decreed through an arbitrary---albeit simple---choice that has in actual fact never been confronted to the verdict of experiment,
but rather be derived from an action principle for point charges meeting all these same requirements, given the absence of any experimental evidence for the
existence of point magnetic charges. Certainly to the author's best knowledge, the latter issue has not been addressed in earnest yet in the literature.
The present work is a contribution in that direction.}

The discussion is organised as follows. Section 2 addresses the structure inherent to the SO(2) dual covariant equations of dual electromagnetism,
in terms of a doubled set of 4-vector gauge potentials which do away with the need of Dirac strings for magnetic monopoles, and whose variational action
in Lagrangian (and Hamiltonian) form is manifestly invariant under space-time Poincar\'e, Lorentz and discrete symmetries,
under SO(2) dual transformations, and under a doubled set of local abelian gauge transformations in the electric and magnetic sectors.
Such or similar considerations may be found in different forms in the (vast) literature on these subjects, dating back to original observations
made in Refs.\cite{Dirac2,Cabibbo} \textcolor{black}{(as well as in Ref.\cite{Barker} and further references therein, for example; see also the reviews in
Refs.\cite{Blago,Singleton,Milton})}.
\textcolor{black}{This doubled gauge potential formulation is} presented herein for the sake of completeness.
Section 3 then considers the coupling to dual electromagnetism of a collection of relativistic
and massive electric and magnetic point charges, to construct a SO(2) dual covariant electrodynamics deriving from an action principle sharing
all the same set of symmetries, thereby identifying the sole possible generalised Lorentz force which proves to involve a single free and real additional
but presently unknown coupling constant channeling interactions between the electric and magnetic sectors of the system which otherwise would be totally
``dark'' to each other. Section 4 then applies these results to a pair of dyons of which one remains static, to ascertain whether given the Lorentz force identified
through the analysis, a quantisation condition for electric and magnetic charges of the Dirac-Schwinger-Zwanziger type would apply for
SO(2) dual electrodynamics, and how to possibly access experimentally its ``dark'' dual magnetic sector through the use of the purely electric sector
which makes up our visible world. Some final considerations are presented in the Conclusions. While an Appendix brings together similar considerations
for reference in the main text, in the case of ordinary Maxwell electrodynamics involving purely electric sources only.

\section{Maxwell Equations of SO(2) Dual Covariant Electromagnetism}
\label{Sect2}

Let us consider thus Maxwell's equations for dual electromagnetism as given in (\ref{eq:DualEM}). It then readily follows that the relativistic wave equations
for the $\vec{E}$ and $\vec{B}$ fields read, whose SO(2) dual covariance is again manifest,
\begin{eqnarray}
\square\,\frac{\vec{E}}{c} &=&
\left(\frac{\partial^2}{\partial(ct)^2}\,-\,\vec{\nabla}^2\right) \frac{\vec{E}}{c} = -\mu_0\vec{\nabla}(c\rho_e)-\mu_0\frac{\partial}{\partial (ct)}\vec{J}_e
-\mu_0\vec{\nabla}\times\vec{J}_m, \nonumber \\
\square\,\vec{B} &=&
\left(\frac{\partial^2}{\partial(ct)^2}\,-\,\vec{\nabla}^2\right) \vec{B} = 
-\mu_0\vec{\nabla}(c\rho_m)-\mu_0\frac{\partial}{\partial (ct)}\vec{J}_m + \mu_0\vec{\nabla}\times\vec{J}_e.
\end{eqnarray}
On the other hand what would be the continuity equation for the conservation of electromagnetic energy density with its flux of energy given by
the Poynting vector in their expressions as they apply for ordinary electromagnetism, is now obtained in the form of,
\begin{equation}
\frac{\partial}{\partial t}\left[\frac{1}{2\mu_0}\left(\left(\frac{\vec{E}}{c}\right)^2+\vec{B}\,^2\right)\right]\,+\,
\vec{\nabla}\cdot\left(\frac{1}{\mu_0}\vec{E}\times\vec{B}\right)=-\vec{J}_e\cdot\vec{E}\,-\,\vec{J}_m\cdot\vec{B},
\end{equation}
a result which points to the necessity of extending at least the expression for the Lorentz force as well, as it should apply in the context of dual electromagnetism.
This issue is addressed in the next Section.

To make Lorentz covariance of the SO(2) dual covariant Maxwell equations manifest, let us introduce the 2-index antisymmetric field strength tensor $F_{\mu\nu}$
defined in terms of the electric and magnetic fields in exactly the same way as in ordinary electromagnetism \cite{Jackson}.
Herein space-time contravariant coordinates
are denoted $x^\mu=(ct,\vec{x}\,)=(ct,x^i)$ ($\mu=0,1,2,3$, $i=1,2,3$), the Minkowski space-time metric is $\eta_{\mu\nu}={\rm diag}\,(+---)$,
while the 4-index totally antisymmetric tensor $\epsilon^{\mu\nu\rho\sigma}$ takes the value $\epsilon^{0123}=+1$. The field strength tensor and its dual are then such
that,
\begin{equation}
F_{\nu\mu}=-F_{\mu\nu},\qquad
{^*\!}F_{\mu\nu} \equiv \frac{1}{2}\epsilon_{\mu\nu\rho\sigma}\,F^{\rho\sigma},\qquad
{^*\!}\left({^*\!}F\right)_{\mu\nu}=-F_{\mu\nu},\qquad
{^*\!}F_{\nu\mu}=-{^*\!}F_{\mu\nu},
\end{equation}
while its entries (all of the same physical dimension) are defined to be,
\begin{equation}
F_{0i}=\frac{E^i}{c},\qquad
F_{ij}=-\epsilon^{ijk}\,B^k,\qquad
{^*\!}F_{0i}=B^i,\qquad
{^*\!}F_{ij}=\epsilon^{ijk}\,\frac{E^k}{c}.
\label{eq:Fmunu}
\end{equation}
Note that the discrete duality transformation which exchanges the electric and magnetic fields now corresponds to the simple map
$F_{\mu\nu}\rightarrow {^*\!}F_{\mu\nu}$ and ${^*\!}F_{\mu\nu}\rightarrow -F_{\mu\nu}={^*\!}\left({^*\!}F_{\mu\nu}\right)$.

Likewise let us combine the charge and current densities in each of the electric and magnetic sectors of charge distributions into the
components of the following conserved 4-vector current densities (all of whose entries are of the same physical dimension),
\begin{equation}
J^\mu_e=(c\rho_e,\vec{J}_e),\qquad
J^\mu_m=(c\rho_m,\vec{J}_m),\qquad
\partial_\mu J^\mu_{e,m}=0.
\end{equation}

As is well known (in the case of ordinary electromagnetism) the two 3-vector and two 3-scalar dual Maxwell equations in (\ref{eq:DualEM}) then combine
into the following two 4-vector Maxwell equations in manifestly Lorentz covariant form, this time with a nonvanishing magnetic source term for the
dual field strength field ${^*\!}F_{\mu\nu}$ as well,
\begin{equation}
\partial^\mu F_{\mu\nu}= \mu_0 J_{e,\nu},\qquad
\partial^\mu\,{^*\!}F_{\mu\nu}=\mu_0 J_{m,\nu}.
\end{equation}
Not only does the space-time covariant notation make the Poincar\'e and Lorentz covariance of dual electromagnetism for each of these
equations separately manifest, with readily identified Lorentz transformations for all coordinates $x^\mu$, fields $F_{\mu\nu}$ and ${^*\!}F_{\mu\nu}$,
and current densities $J^\mu_{e,m}$, but in the presence of the magnetic current density $J^\mu_m$ the SO(2) dual covariance property of this pair of
Lorentz covariant equations also becomes manifest, while mixing the electric and magnetic matter sources for the fields $F_{\mu\nu}$ and ${^*\!}F_{\mu\nu}$
according to the relations,
\begin{equation}
\left(\begin{array}{c}
F_{\mu\nu}' \\
{^*\!}F_{\mu\nu}'
\end{array}\right)=
\left(\begin{array}{c c}
\cos\theta & \sin\theta \\
-\sin\theta & \cos\theta
\end{array}\right)\cdot
\left(\begin{array}{c}
F_{\mu\nu} \\
{^*\!}F_{\mu\nu}
\end{array}\right),\quad
\left(\begin{array}{c}
{J^\mu_e}' \\
{J^\mu_m}'
\end{array}\right)=
\left(\begin{array}{c c}
\cos\theta & \sin\theta \\
-\sin\theta & \cos\theta
\end{array}\right)\cdot
\left(\begin{array}{c}
J^\mu_e \\
J^\mu_m
\end{array}\right).
\end{equation}

To proceed with the analysis of these Maxwell equations in this manifest Lorentz and SO(2) dual covariant form, let us consider the 2-form
$F=\frac{1}{2}dx^\mu\wedge dx^\nu\,F_{\mu\nu}$, and in particular its Hodge decomposition\footnote{The parallel analysis for ordinary
electromagnetism is addressed in the Appendix.}. Then simply on account of the indisputable verdict of mathematical considerations one may conclude that
the field strengths $F_{\mu\nu}$ and ${^*\!}F_{\mu\nu}$ may always be given the following two representations in terms of the curls of
two independent 4-vector fields $A_\mu$ and $C_\mu$, namely a doubled set of 4-vector gauge potentials,
\begin{eqnarray}
F_{\mu\nu} &=& \left(\partial_\mu A_\nu  - \partial_\nu A_\mu\right) \, - \, \frac{1}{2}\epsilon_{\mu\nu\rho\sigma}\left(\partial^\rho C^\sigma - \partial^\sigma C^\rho\right)
\equiv A_{\mu\nu}-\frac{1}{2}\epsilon_{\mu\nu\rho\sigma}\,C^{\rho\sigma}=A_{\mu\nu}-{^*}C_{\mu\nu}, \nonumber \\
{^*\!}F_{\mu\nu} &=& \left(\partial_\mu C_\nu  - \partial_\nu C_\mu\right) \, + \, \frac{1}{2}\epsilon_{\mu\nu\rho\sigma}\left(\partial^\rho A^\sigma - \partial^\sigma
A^\rho\right) \equiv C_{\mu\nu}+\frac{1}{2}\epsilon_{\mu\nu\rho\sigma}\, A^{\rho\sigma}=C_{\mu\nu}+{^*\!}A_{\mu\nu},
\end{eqnarray}
with the convenient short-hand notations
\begin{equation}
A_{\mu\nu}\equiv \partial_\mu A_\nu  - \partial_\nu A_\mu,\qquad
C_{\mu\nu}\equiv \partial_\mu C_\nu  - \partial_\nu C_\mu,
\end{equation}
while the 4-vector potential 1-forms $A=dx^\mu\,A_\mu$ and $C=dx^\mu\,C_\mu$ are each defined up to an exact 1-form, namely up to some gauge transformation
in the form of
\begin{equation}
A'_\mu=A_\mu + \partial_\mu \Lambda_A,\qquad
C'_\mu=C_\mu + \partial_\mu \Lambda_C,
\end{equation}
$\Lambda_A(x^\mu)$ and $\Lambda_C(x^\mu)$ being two arbitrary scalar fields. Under the assumption that these fields vanish at infinity this gauge freedom
allows to require that the gauge potentials $A_\mu$ and $C_\mu$ are restricted to obey, for instance, the Lorenz gauge fixing condition,
\begin{equation}
\partial^\mu A_\mu=0,\qquad
\partial^\mu C_\mu=0
\end{equation}
(gauge transformations preserving these Lorenz gauge conditions must be such that $\square\Lambda_A=0$ and $\square\Lambda_C=0$,
with of course $\square\equiv \partial^\mu\partial_\mu$).

Obviously the action of SO(2) duality transformations mixes these two gauge potentials once again according to the SO(2) rotation 
between the two electric and magnetic sectors of dual electromagnetism, namely,
\begin{equation}
\left(\begin{array}{c}
A_\mu^{'} \\
C_\mu^{\,'}
\end{array}\right)=
\left(\begin{array}{c c}
\cos\theta & \sin\theta \\
-\sin\theta & \cos\theta
\end{array}\right)\cdot
\left(\begin{array}{c}
A_\mu \\
C_\mu
\end{array}\right),\qquad
\left(\begin{array}{c}
A_{\mu\nu}^{'} \\
C_{\mu\nu}^{\,'}
\end{array}\right)=
\left(\begin{array}{c c}
\cos\theta & \sin\theta \\
-\sin\theta & \cos\theta
\end{array}\right)\cdot
\left(\begin{array}{c}
A_{\mu\nu} \\
C_{\mu\nu}
\end{array}\right).
\end{equation}

Since (on account of the Bianchi identity) $\partial^\mu\,{^*\!}A_{\mu\nu}=0$ and $\partial^\mu\,{^*}C_{\mu\nu}=0$, the dual Maxwell equations reduce to,
\begin{equation}
\partial^\mu F_{\mu\nu}=\partial^\mu A_{\mu\nu}=\mu_0 J_{e,\nu},\qquad
\partial^\mu{^*\!}F_{\mu\nu}=\partial^\mu C_{\mu\nu}=\mu_0 J_{m,\nu},
\end{equation}
so that in terms of these gauge potentials the above dual Maxwell equations are then of second-order in space-time derivatives,
\begin{equation}
\square A_\mu - \partial_\mu\left(\partial^\nu A_\nu\right)=\mu_0\,J_{e,\mu},\qquad
\square C_\mu - \partial_\mu\left(\partial^\nu C_\nu\right) = \mu_0 J_{m,\mu},\qquad
\end{equation}
which in the Lorenz gauge thus reduce simply to
\begin{equation}
\square A_\mu = \mu_0 J_{e,\mu},\qquad
\square C_\mu=\mu_0 J_{m,\mu}.
\end{equation}
Up to the doubled gauge transformations in $\Lambda_A$ and $\Lambda_C$, the two gauge potentials $A_\mu$ and $C_\mu$ are thus determined separately
by the two conserved electric and magnetic 4-current sources $J^\mu_e$ and $J^\mu_m$, respectively. While Lorentz transformations map each of these two sectors into themselves separately, these two sectors of dual electromagnetism get mixed into one another in a continuous fashion
under arbitrary SO(2) dual transformations parametrised by the angle $\theta$.

By considering the components of the 4-vector gauge potentials in the form
\begin{equation}
A^\mu=\left(\frac{\Phi_A}{c},\vec{A}\right),\qquad
C^\mu=\left(\frac{\Phi_C}{c},\vec{C}\right),
\end{equation}
and defining ``electric'' and ``magnetic'' fields in each of the electric and magnetic sectors of dual electromagnetism according to the expressions,
\begin{equation}
\frac{E^i_e}{c}=A_{0i},\qquad B^i_e={^*\!}A_{0i},\qquad
\frac{E^i_m}{c}=C_{0i},\qquad B^i_m={^*}C_{0i},
\end{equation}
one finds,
\begin{equation}
\vec{E}_e=-\vec{\nabla}\Phi_A-\partial_t\vec{A},\qquad
\vec{B}_e=\vec{\nabla}\times\vec{A},\qquad
\vec{E}_m=-\vec{\nabla}\Phi_C-\partial_t\vec{C},\qquad
\vec{B}_m=\vec{\nabla}\times\vec{C}.
\end{equation}
Thus finally the original electric and magnetic fields $\vec{E}$ and $\vec{B}$ are decomposed as,
\begin{equation}
\frac{\vec{E}}{c}=\frac{\vec{E}_e}{c}-\vec{B}_m=-\vec{\nabla}\frac{\Phi_A}{c}-\frac{\partial}{\partial(ct)}\vec{A}-\vec{\nabla}\times\vec{C},\qquad
\vec{B}=\frac{\vec{E}_m}{c}+\vec{B}_e=-\vec{\nabla}\frac{\Phi_C}{c}-\frac{\partial}{\partial(ct)}\vec{C}+\vec{\nabla}\times\vec{A}.
\label{eq:E-B-em}
\end{equation}
Furthermore in 3-vector and 3-scalar form the dual Maxwell equations read, in the electric sector,
\begin{equation}
\vec{\nabla}\cdot\frac{\vec{E}_e}{c}=\mu_0(c\rho_e),\quad
\vec{\nabla}\times\frac{\vec{E}_e}{c}+\frac{\partial}{\partial(ct)}\vec{B}_e=\vec{0},\quad
\vec{\nabla}\cdot\vec{B}_e=0,\quad
\vec{\nabla}\times\vec{B}_e-\frac{\partial}{\partial(ct)}\frac{\vec{E}_e}{c}=\mu_0\,\vec{J}_e,
\end{equation}
and in the magnetic sector,
\begin{equation}
\vec{\nabla}\cdot\frac{\vec{E}_m}{c}=\mu_0(c\rho_m),\quad
\vec{\nabla}\times\frac{\vec{E}_m}{c}+\frac{\partial}{\partial(ct)}\vec{B}_m=\vec{0},\quad
\vec{\nabla}\cdot\vec{B}_m=0,\quad
\vec{\nabla}\times\vec{B}_m-\frac{\partial}{\partial(ct)}\frac{\vec{E}_m}{c}=\mu_0\,\vec{J}_m.
\end{equation}
It may readily be checked that by using the above expressions for $\vec{E}$ and $\vec{B}$ in terms of the fields $\vec{E}_{e,m}$ and $\vec{B}_{e,m}$,
the dual Maxwell equations in (\ref{eq:DualEM}) are indeed recovered, as it should of course.

Obviously (from the knowledge for ordinary electromagnetism---see the Appendix) all these equations of motion derive from the following action principle
in the gauge fields $A_\mu$ and $C_\mu$, $J^\mu_{e,m}$ being external sources,
\begin{eqnarray}
&&S[A^\mu,C^\mu;J^\mu_e,J^\mu_m]= \nonumber \\
&=&\int d^4x^\mu\left\{-\frac{1}{4\mu_0 c}\left(\partial_\mu A_\nu - \partial_\nu A_\mu\right)^2
-\frac{1}{4\mu_0 c}\left(\partial_\mu C_\nu - \partial_\nu C_\mu\right)^2-\frac{1}{c}A_\mu J^\mu_e - \frac{1}{c} C_\mu J^\mu_m\right\},
\label{eq:LDual}
\end{eqnarray}
where this Lagrangian density is defined up to a total space-time surface term. On account of the conservation of each of the electric and magnetic
current densities, $\partial_\mu J^\mu_{e,m}=0$, clearly this action is invariant (up to such a surface term)
under the gauge transformations in $\Lambda_A$ and $\Lambda_C$ of the gauge fields $A_\mu$ and $C_\mu$.
Its invariance both under space-time Poincar\'e and Lorentz transformations, as well as SO(2) dual transformations,
is also manifest. The corresponding first-order Hamiltonian action sharing precisely these same symmetries explicitly is given as,
\begin{eqnarray}
&&S[A^\mu,\phi^{\mu\nu}_e,C^\mu,\phi^{\mu\nu}_m;J^\mu_e,J^\mu_m] = \nonumber \\
&=& \int d^4x^\mu\left\{\frac{1}{4\mu_0 c}\phi_{e,\mu\nu}\phi^{\mu\nu}_e-\frac{1}{2\mu_0 c}\phi_{e,\mu\nu}\left(\partial^\mu A^\nu - \partial^\nu A^\mu\right)
-\frac{1}{c} A_\mu J^\mu_e \right. \nonumber \\
&&\left. \qquad\quad\  +\frac{1}{4\mu_0 c}\phi_{m,\mu\nu}\phi^{\mu\nu}_m-\frac{1}{2\mu_0 c}\phi_{m,\mu\nu}\left(\partial^\mu C^\nu - \partial^\nu C^\mu\right)
-\frac{1}{c}C_\mu J^\mu_m\right\},
\end{eqnarray}
which implies the following Lorentz and SO(2) dual covariant Hamiltonian equations of motion for the real gaussian 2-index antisymmetric
tensor fields $\phi^{\mu\nu}_{e,m}$ (see also the Appendix for further comments),
\begin{equation}
\phi_{e,\mu\nu}=A_{\mu\nu}=\partial_\mu A_\nu - \partial_\nu A_\mu,\qquad
\phi_{m,\mu\nu}=C_{\mu\nu}=\partial_\mu C_\nu - \partial_\nu C_\mu,
\end{equation}
thereby reproducing the original Lagrangian equations of motion for the fields $A_\mu$ and $C_\mu$,
\begin{equation}
\partial^\mu A_{\mu\nu} = \mu_0 J_{e,\nu},\qquad
\partial^\mu C_{\mu\nu} = \mu_0 J_{m,\nu},
\end{equation}
since one also has the following Hamiltonian equations of motion for these latter fields,
\begin{equation}
\partial^\mu \phi_{e,\mu\nu}=\mu_0 J_{e,\nu},\qquad
\partial^\mu \phi_{m,\mu\nu}=\mu_0 J_{m,\nu}.
\end{equation}

Finally, since the following two continuity equations apply in each of the electric and magnetic sectors,
\begin{eqnarray}
\partial_t\left[\frac{1}{2\mu_0}\left(\left(\frac{\vec{E}_e}{c}\right)^2+\vec{B}_e\,^2\right)\right] + \vec{\nabla}\cdot\left(\frac{1}{\mu_0}\vec{E}_e\times\vec{B}_e\right)
&=& - \vec{J}_e\cdot\vec{E}_e, \nonumber \\
\partial_t\left[\frac{1}{2\mu_0}\left(\left(\frac{\vec{E}_m}{c}\right)^2+\vec{B}_m\,^2\right)\right] + \vec{\nabla}\cdot\left(\frac{1}{\mu_0}\vec{E}_m\times\vec{B}_m\right)
&=& - \vec{J}_m\cdot\vec{E}_m,
\end{eqnarray}
the total dual electromagnetic energy density and total dual Poynting vector for the flux of this total energy density are given by, respectively,
\begin{eqnarray}
u &=& \frac{1}{2\mu_0}\left(\left(\frac{\vec{E}_e}{c}\right)^2+\vec{B}_e\,^2\right) + \frac{1}{2\mu_0}\left(\left(\frac{\vec{E}_m}{c}\right)^2+\vec{B}_m\,^2\right),
\nonumber \\
\vec{S} &=& \frac{1}{\mu_0}\vec{E}_e\times\vec{B}_e + \frac{1}{\mu_0}\vec{E}_m\times\vec{B}_m,
\label{eq:Dual-u}
\end{eqnarray}
thus obeying the continuity equation for the conservation of dual electromagnetic energy,
\begin{equation}
\partial_t u + \vec{\nabla}\cdot\vec{S}=- \vec{J}_e\cdot\vec{E}_e -  \vec{J}_m\cdot\vec{E}_m.
\end{equation}
Note how these two quantities, $u$ and $\vec{S}$, are indeed SO(2) dual invariant.
It may easily be checked that these latter quantities do not correspond to those of ordinary electromagnetism,
namely $((\vec{E}/c)^2+\vec{B}^2)/(2\mu_0)$ and $\vec{E}\times\vec{B}/\mu_0$, respectively (even though the latter two are SO(2) dual invariant as well),
once the expressions for $\vec{E}$ and $\vec{B}$ 
in terms of $\vec{E}_{e,m}$ and $\vec{B}_{e,m}$ are substituted. Nonetheless they of course reduce to those in (\ref{eq:Dual-u})
once the magnetic sector is suppressed, namely when $J^\mu_m=0$, or $\rho_m=0$ and $\vec{J}_m=\vec{0}$, so that $C^\mu$ is pure gauge,
or $C^\mu=0$ up to a gauge transformation, which implies $C_{\mu\nu}=0$.

\begin{table}
\begin{center}
\begin{tabular}{|c|c|c||c|c|c|}
\hline
 & P & T & & P & T \\
 \hline
 $\rho_e$ & $+$ & $+$ & $\rho_m$ & $-$ & $-$ \\
 $\vec{J}_e$ & polar & $-$ & $\vec{J}_m$ & axial & $+$ \\
 $A^0$ & $+$ & $+$ & $C^0$ & $-$ & $-$ \\
 $\vec{A}$ & polar & $-$ & $\vec{C}$ & axial & $+$ \\
 $\vec{E}_e$ & polar & $+$ & $\vec{E}_m$ & axial & $-$ \\
 $\vec{B}_e$ & axial & $-$ & $\vec{B}_m$ & polar & $+$ \\
 $\vec{E}$ & polar & $+$ & $\vec{B}$ & axial & $-$ \\
 \hline
\end{tabular}
\caption[]{Parity (P) and time reversal (T) property assignments for electromagnetic quantities.}
\label{Table1}
\end{center}
\end{table}

Besides the manifest continuous Poincar\'e and SO(2) dual symmetries considered so far, it may easily be confirmed that the classical action (\ref{eq:LDual})
is also invariant under discrete parity (P) and time reversal (T) transformations, given the P and T assignment properties listed in Table~\ref{Table1} for all
electromagnetic quantities. \textcolor{black}{Incidentally note that the SO(2) duality rotation angle $\theta$ is also a pseudoscalar quantity under
parity (P) \cite{Singleton}, and is odd under time reversal (T).}

One last feature of SO(2) invariant dual electromagnetism certainly still needs to be emphasized as well. In contradistinction to ordinary electromagnetism with its
sole 4-vector gauge potential $A^\mu=(\Phi_A/c,\vec{A})$ leading to the nonlocal Dirac string singularity for the vector potential carried along
by any point magnetic monopole, in SO(2) dual electromagnetism with its two independent 4-vector gauge potentials $A^\mu$ and $C^\mu$,
magnetic monopoles come along without any strings attached. The sole singularities of these gauge potentials are the usual ones of Coulomb type
characteristic of any point charge in 3-Euclidean space, be it electric or now magnetic.
In SO(2) dual electromagnetism the total $\vec{B}$ field of a magnetic monopole is rather determined directly from a Coulomb-like
scalar potential---as opposed to a vector potential---in the magnetic sector.

More explicitly, let us consider sources $J^\mu_{e,m}$ such that only the component $J^0_m$ is nonvanishing and corresponds to a point magnetic monopole
of charge $g$ positioned at the origin of the inertial frame in use, namely,
\begin{equation}
\rho_e=0,\qquad
\vec{J}_e=\vec{0};\qquad
\rho_m(ct,\vec{x})=g\,\delta^{(3)}(\vec{x}\,),\qquad
\vec{J}_m=\vec{0}.
\end{equation}
Given the Maxwell equations in the Lorenz gauge, namely $\square A^\mu=\mu_0 J^\mu_e$ as well as $\square C^\mu=\mu_0 J^\mu_m$,
a static solution is such that $A^\mu=0$ in the electric sector, while in the magnetic sector the vector potential $\vec{C}$ vanishes identically as well,
$\vec{C}=\vec{0}$, leaving as only nontrivial potential the magnetic scalar potential $\Phi_C$ obeying the Poisson equation,
\begin{equation}
-\vec{\nabla}^2\Phi_C(\vec{x})=\mu_0 c^2 g\,\delta^{(3)}(\vec{x}\,)=\frac{1}{\epsilon_0}\,g\,\delta^{(3)}(\vec{x}\,),
\end{equation}
with the ready Coulomb-like solution,
\begin{equation}
\Phi_C(\vec{x}\,)=\frac{g}{4\pi\epsilon_0}\cdot\frac{1}{|\vec{x}\,|},\qquad
\vec{E}_m(\vec{x}\,)=-\vec{\nabla}\Phi_C(\vec{x}\,)=\frac{g}{4\pi\epsilon_0}\cdot\frac{\vec{x}}{|\vec{x}\,|^3},
\end{equation}
while the remaining electric and magnetic fields in the electric and magnetic sectors vanish identically,
$\vec{E}_e=\vec{0}$, $\vec{B}_e=\vec{0}$, and $\vec{B}_m=\vec{0}$. Consequently the total electric and magnetic fields, $\vec{E}$ and $\vec{B}$,
are indeed those of an electrically neutral point magnetic monopole of charge $g$ (given our choice of units),
\begin{equation}
\vec{E}=\vec{0},\qquad
\vec{B}(\vec{x}\,)=\frac{\vec{E}_m(\vec{x}\,)}{c}=\frac{g}{4\pi\epsilon_0 c}\cdot\frac{\vec{x}}{|\vec{x}\,|^3}=\frac{\mu_0 c}{4\pi}\,g\cdot\frac{\vec{x}}{|\vec{x}\,|^3},
\end{equation}
while none of the associated gauge potentials, whether scalar or vector in the electric and magnetic sectors, displays any nonlocal string singularity of the Dirac type.

\section{SO(2) Dual Covariant Electrodynamics of Point Charges}
\label{Sect3}

Let us now address the coupling to SO(2) dual invariant electromagnetism of a collection ($s=1,2,\cdots,N$) of $N$ electrically and magnetically charged
relativistic point particles\footnote{The coupling to similarly charged relativistic fields is to be considered elsewhere.}, of masses $m_s$,
electric charges $q_s$ and magnetic charges $g_s$, each following its space-time world-line $x^\mu_{(s)}(\tau_s)$ with world-line parametrisation $\tau_s$,
such that relative to the inertial frame in use one has $x^0_{(s)}(\tau_s)=ct$ (this relation determines the change of variable $\tau_s(t) \leftrightarrow t(\tau_s)$
for each world-line. See also the Appendix).

However, in order to identify the possible SO(2) dual invariant generalisation of the ordinary Lorentz force for purely electrically charged point masses, and for
notational simplicity, first the case of a single such particle is to be considered, with mass $m$, charges $q$ and $g$, and world-line $x^\mu(\tau)$.
As is well known (see the Appendix) the ordinary Lorentz force equation (\ref{eq:Lorentz}) is equivalent to the following manifestly Poincar\'e and Lorentz
covariant expression for a massive relativistic point particle,
\begin{equation}
\frac{d}{d\tau}\left(\frac{mc}{\sqrt{(dx^\mu/d\tau)^2}}\,\frac{dx_\mu(\tau)}{d\tau}\right)=q\,F_{\mu\nu}(x^\mu(\tau))\,\frac{dx^\nu(\tau)}{d\tau},
\label{eq:Lorentz-ordi}
\end{equation}
where in this expression the world-line parametrisation in $\tau$ is specifically the world-line proper-time coordinate for that particle's trajectory.

In view of the SO(2) dual transformations of $F_{\mu\nu}$, ${^*\!}F_{\mu\nu}$ and of the charges $(q,g)$, a ready suggestion for a SO(2) dual invariant
generalisation of this Lorentz force, in Lorentz covariant form, is obviously,
\begin{equation}
\frac{d}{d\tau}\left(\frac{mc}{\sqrt{(dx^\mu/d\tau)^2}}\,\frac{dx_\mu(\tau)}{d\tau}\right)=\left(q\,F_{\mu\nu}(x^\mu(\tau))
\,+\,g\,{^*\!}F_{\mu\nu}(x^\mu(\tau))\right)\,\frac{dx^\nu(\tau)}{d\tau},
\label{eq:Lorentz-naive}
\end{equation}
which may be seen to indeed correspond to the ready suggestion pointed out in (\ref{eq:generLorentz}) in the form of,
\begin{equation}
\frac{d}{dt}\vec{p}=qc\left(\frac{\vec{E}}{c}+\frac{\vec{v}}{c}\times\vec{B}\right)\,+\,
gc\left(\vec{B}-\frac{\vec{v}}{c}\times\frac{\vec{E}}{c}\right).
\label{eq:Lorentz-naive2}
\end{equation}
However, would such a purely theoretical (so far) choice for a Lorentz force extended to magnetic charges, albeit a SO(2) dual invariant choice, derive
from a variational principle? And if yes, which one? In addition, would this option then be the only SO(2) dual invariant generalised Lorentz force possible
deriving from an action principle?

Incidentally let us point out that one has,
\begin{equation}
q\, F_{\mu\nu} + g\, {^*\!}F_{\mu\nu}=\left(q\, A_{\mu\nu} + g\, C_{\mu\nu}\right) + \left(g\,{^*\!}A_{\mu\nu} - q\,{^*}C_{\mu\nu}\right),
\label{eq:FdualF}
\end{equation}
which includes contributions specifically of the dual field strengths in the electric and magnetic sectors, ${^*\!}A_{\mu\nu}$ and ${^*}C_{\mu\nu}$.
In particular given the electric and magnetic field strengths and their duals, and the charges $(q,g)$, the only two possible SO(2) dual invariants
linear in each of these two types of physical quantities, namely fields and charges, are precisely the two combinations,
\begin{equation}
q\, A_{\mu\nu} + g\, C_{\mu\nu},\qquad
g\,{^*\!}A_{\mu\nu} - q\,{^*}C_{\mu\nu}.
\label{eq:blocks}
\end{equation}
These are thus the two building blocks to be used hereafter in the search of a suitable SO(2) dual invariant generalised Lorentz force.

In order to address these questions, let consider the action for such a (massive) relativistic point particle in the following general form,
\begin{equation}
S[x^\mu,\lambda]=\int d\tau L\left(x^\mu(\tau),\frac{d x^\mu(\tau)}{d\tau};\lambda(\tau)\right),
\end{equation}
where the Lagrange function comprises a purely kinetic contribution (the sum of the first two terms on the r.h.s.~of the expression hereafter)
to which is added an interaction term, $L_{int}$, which couples in a manifestly SO(2) dual invariant way the particle to the electromagnetic fields
and gauge potentials of dual electromagnetism,
\begin{equation}
L\left(x^\mu,\frac{dx^\mu}{d\tau};\lambda\right)=-\frac{1}{2}\lambda^{-1}c\left(\frac{dx^\mu}{d\tau}\right)^2-\frac{1}{2}\lambda m^2c
\,+\,L_{int}\left(x^\mu,\frac{dx^\mu}{d\tau}\right).
\end{equation}
In the kinetic contribution, $\lambda(\tau)>0$ is a world-line ein-bein, playing the role as well of a Lagrange multiplier for the generator (and first-class constraint)
of world-line diffeomorphisms in $\tau$ ($L_{int}$ does not involve $\lambda$). The variational equation for $\lambda(\tau)$ implies that
$\lambda(\tau)=\left(d x^\mu(\tau)/d\tau\right)^{1/2}/m$, which corresponds to the proper-time world-line parametrisation
in the case of a massive relativistic point particle (see the Appendix).

Consider first the usual choice for such a coupling, involving some vector field $W_{(0)}(x^\mu)$
and its field strength $W_{(0)\mu\nu}\equiv\partial_\mu W_{(0)\nu} - \partial_\nu W_{(0)\mu}$,
thus with the following interaction Lagrangian,
\begin{equation}
L^{(0)}_{int}\left(x^\mu,\frac{dx^\mu}{d\tau}\right)=-\frac{dx^\mu}{d\tau}\cdot W_{(0)\mu}(x^\mu),\qquad
d\tau\,L^{(0)}_{int}\left(x^\mu,\frac{dx^\mu}{d\tau}\right) = -dx^\mu \cdot W_{(0)\mu}(x^\mu).
\end{equation}
In the proper-time world-line parametrisation the corresponding equation of motion reads,
\begin{equation}
\frac{d}{d\tau}\left(\frac{mc}{\sqrt{\left(\frac{d x^\mu}{d\tau}\right)^2}}\,\frac{d x_\mu}{d\tau}\right)=W_{(0)\mu\nu}\cdot\frac{dx^\nu}{d\tau}.
\end{equation}
Consequently such an interaction coupling cannot produce a Lorentz force which includes a contribution from the field strength ${^*}W_{(0)\mu\nu}$
which is dual to $W_{(0)\mu\nu}$. In other words with only the two 4-vector gauge potentials $A^\mu$ and $C^\mu$ available for dual electromagnetism,
couplings of the above (and usual) type cannot lead to a Lorentz force which involves as well a coupling to the duals ${^*\!}A_{\mu\nu}$
and ${^*}C_{\mu\nu}$ and not only $A_{\mu\nu}$ and $C_{\mu\nu}$, as the theoretical choice (\ref{eq:Lorentz-naive}) would require.

In order to look for possible alternative choices of interaction couplings, let us note that by integration by parts in $\tau$ the above usual coupling
may also be expressed as follows, up to a total $\tau$-derivative term,
\begin{equation}
dx^\mu\cdot W_{(0)\mu}(x^\mu)=d\left(x^\mu\cdot W_{(0)\mu}(x^\mu)\right)\,-\,dx^\mu\,x^\nu\partial_\mu W_{(0)\nu}(x^\mu),
\end{equation}
or equivalently,
\begin{equation}
d\tau\,\frac{dx^\mu}{d\tau}\cdot W_{(0)\mu}(x^\mu)=d\tau\,\frac{d}{d\tau}\left(x^\mu\cdot W_{(0)\mu}(x^\mu)\right)
\,-\,d\tau\,\frac{dx^\mu}{d\tau}\,x^\nu\partial_\mu W_{(0)\nu}(x^\mu).
\end{equation}
In this partially integrated form, the interaction coupling is linear in $dx^\mu/d\tau$, in $x^\mu$, and in the 4-vector field $W_{(0)\mu}(x^\mu)$.
On the other hand, the dependency of the Lorentz force needs to be at most linear in $dx^\mu/d\tau$ as well as in (a field derived from) $W_{(0)\mu}$.
This requires that the interaction coupling may depend at most linearly on $dx^\mu/d\tau$ and on the 4-vector field.

Given this remark, the most general choice possible for the interaction coupling may involve up to four different 4-vector fields, $W_{(\alpha)\mu}(x^\mu)$
($\alpha=0,1,2,3$), with an interaction Lagrangian of the following form (of course the fields $W_{(0)\mu}$ and $W_{(1)\mu}$ will contribute jointly and on equal
terms to the Lorentz force equation of motion),
\begin{equation}
L_{int}\left(x^\mu,\frac{d x^\mu}{d\tau}\right)=
-\frac{d x^\mu}{d\tau}\left(W_{(0)\mu}(x^\mu)\,-\,x^\nu\left(\partial_\mu W_{(1)\nu}(x^\mu)+\partial_\nu W_{(2)\mu}(x^\mu)
+\epsilon_{\mu\nu\lambda\eta}\partial^\lambda W^\eta_{(3)}(x^\mu)\right)\right).
\end{equation}
While the space-time Lorentz invariance of this action is manifest,
it may readily be checked, up to a total $\tau$-derivative, that this action is as well invariant under abelian gauge redefinitions
of these background vector fields, namely,
\begin{equation}
W'_{(\alpha)\mu}(x^\mu)=W_{(\alpha)\mu}(x^\mu)+\partial_\mu \Lambda_{(\alpha)}(x^\mu),\qquad \alpha=0,1,2,3,
\end{equation}
where $\Lambda_{(\alpha)}(x^\mu)$ are arbitrary scalar fields. Indeed one simply has,
\begin{equation}
\frac{dx^\mu}{d\tau}\partial_\mu\Lambda_{(\alpha)}=\frac{d}{d\tau}\Lambda_{(\alpha)},\qquad
\frac{d x^\mu}{d\tau} x^\nu \partial_\mu \partial_\nu \Lambda_{(\alpha)}=\frac{d}{d\tau}\left(x^\nu \partial_\nu \Lambda_{(\alpha)} - \Lambda_{(\alpha)}\right).
\end{equation}
Consequently the Lorentz force equation of motion will dependent explicitly only on the associated field strengths, defined as,
\begin{equation}
W_{(\alpha)\mu\nu}\equiv \partial_\mu W_{(\alpha)\nu} - \partial_\nu W_{(\alpha)\mu},\qquad
{^*}W_{(\alpha)\mu\nu}\equiv \frac{1}{2}\epsilon_{\mu\nu\rho\sigma} W^{\rho\sigma}_{(\alpha)},\qquad
\alpha=0,1,2,3.
\end{equation}

Working out the Lorentz force equation, one finds indeed, again in the proper-time world-line parametrisation,
\begin{eqnarray}
\frac{d}{d\tau}\left(\frac{mc}{\sqrt{\left(\frac{d x^\mu}{d\tau}\right)^2}}\frac{d x_\mu}{d\tau}\right) &=& 
\quad \frac{d x^\nu}{d\tau}\left(W_{(0)\mu\nu} + W_{(1)\mu\nu}-W_{(2)\mu\nu}+\epsilon_{\mu\nu\lambda\eta} W^{\lambda\eta}_{(3)}\right) \nonumber \\
&& - \frac{d x^\nu}{d\tau} x^\rho\left(\partial_\rho W_{(2)\mu\nu}+\epsilon_{\nu\rho\lambda\eta}\partial_\mu\partial^\lambda W^\eta_{(3)}
-\epsilon_{\mu\rho\lambda\eta}\partial_\nu\partial^\lambda W^\eta_{(3)}\right) \nonumber \\
&=& \quad \left(W_{(0)\mu\nu}+W_{(1)\mu\nu}-W_{(2)\mu\nu}+2\,{^*}W_{(3)\mu\nu}\right)\frac{d x^\nu}{d\tau} \nonumber \\
&& - x^\rho\left(\partial_\rho W_{(2)\mu\nu} + \partial_\mu{^*}W_{(3)\nu\rho} - \partial_\nu {^*}W_{(3)\mu\rho}\right)\frac{d x^\nu}{d\tau}.
\end{eqnarray}
Note how the coupling in $W_{(3)\mu}(x^\mu)$ now enables a Lorentz force which involves dual field strengths as well.

Considering now the requirement of SO(2) dual invariant couplings, the fields $W_{(\alpha)\mu}$ are necessarily obtained through linear
combinations of the only two available SO(2) dual invariant building blocks in (\ref{eq:blocks}), with real and dimensionless
coefficients $\alpha_1$, $\alpha_2$, $\alpha_3$,
$\beta_1$, $\beta_2$ and $\beta_3$, in the form of,
\begin{eqnarray}
W_{(0)\mu}+W_{(1)\mu} &=& \alpha_1\left(q A_\mu+g C_\mu\right) + \beta_1 \left(g A_\mu - q C_\mu\right), \nonumber \\
W_{(2)\mu} &=& \alpha_2 \left(q A_\mu + g C_\mu\right) + \beta_2 \left(g A_\mu - q C_\mu\right), \nonumber \\
W_{(3)\mu} &=& \frac{1}{2}\alpha_3 \left(q A_\mu + g C_\mu\right) + \frac{1}{2}\beta_3 \left(g A_\mu - q C_\mu\right).
\end{eqnarray}

Finally, imposing the requirement which is directly motivated by well established experimental facts, namely  
that in the absence of any magnetic charges and currents---so that $g=0$ and $J^\mu_m=0$, thus as well that $C^\mu=0$---the generalised
Lorentz force equation of motion established above need to reduce back to that of ordinary electrodynamics  in (\ref{eq:Lorentz-ordi}),
enforces the following restrictions on the possible values for these real coefficients,
\begin{equation}
\alpha_1=1,\qquad
\beta_1=0,\qquad
\alpha_2=0,\qquad
\beta_2=0,\qquad
\alpha_3=0,
\end{equation}
while leaving only $\beta_3\in\mathbb{R}$ as a totally free and independent parameter at this classical level.

Correspondingly, the Lorentz force equation of motion, in manifest Lorentz covariant form and for the proper-time world-line parametrisation,
is then given by the expression,
\begin{eqnarray}
\frac{d}{d\tau}\left(\frac{mc}{\sqrt{\left(\frac{d x^\mu}{d\tau}\right)^2}}\frac{d x_\mu}{d\tau}\right) &=& 
\left[q\, F_{\mu\nu}+g\, {^*\!}F_{\mu\nu}+(\beta_3-1)\left(g\, {^*\!}A_{\mu\nu}-q\,{^*}C_{\mu\nu}\right)\right]\frac{d x^\nu}{d\tau} \nonumber \\
&& -\frac{1}{2}\beta_3\left[g\,\left(\partial_\mu{^*\!}A_{\nu\rho} - \partial_\nu{^*\!}A_{\mu\rho}\right)
- q\, \left(\partial_\mu{^*}C_{\nu\rho}-\partial_\nu{^*}C_{\mu\rho}\right)\right] x^\rho\frac{dx^\nu}{d\tau} \nonumber \\
&=& \left[\left(q\, A_{\mu\nu}+g\, C_{\mu\nu}\right)+\beta_3\left(g\,{^*\!}A_{\mu\nu}-q\, {^*}C_{\mu\nu}\right)\right]\frac{d x^\nu}{d\tau} \nonumber \\
&& -\frac{1}{2}\beta_3\left[g\,\left(\partial_\mu{^*\!}A_{\nu\rho} - \partial_\nu{^*\!}A_{\mu\rho}\right)
- q\, \left(\partial_\mu{^*\!}C_{\nu\rho}-\partial_\nu{^*}C_{\mu\rho}\right)\right] x^\rho\frac{dx^\nu}{d\tau}.
\label{eq:Lorentz-final}
\end{eqnarray}

Incidentally, note that this conclusion also establishes that what would have been a purely theoretically chosen Lorentz force in the form of (\ref{eq:Lorentz-naive})
indeed cannot derive from some variational principle\textcolor{black}{\cite{Rohr}}. At the classical level and requiring an action principle,
the sole freedom afforded for the coupling of electric and magnetic point charges
to dual electromagnetism in a manifestly Poincar\'e, Lorentz and SO(2) dual invariant way, which reduces to the correct and experimentally established and confirmed
formulation of Maxwell's electrodynamics in the absence of magnetic charges and currents, is encoded in the sole real parameter $\beta_3$.
This coupling parametrises the strength of the direct coupling of the particle to the dual field strength of the second SO(2) dual invariant combination
of the dual vector potentials $A^\mu$ and $C^\mu$, namely $(g A_\mu - q C_\mu)$. While the coupling to the first SO(2) dual invariant
combination, namely $(q A_\mu + g C_\mu)$ and its field strength, is through the usual type of coupling with the standard normalisation which is fixed
from experiment and the ordinary Lorentz force.

In explicit form, the dynamics of the massive relativistic dyon of charges $(q,g)$ thus derives from the following
Lagrange function, which manifestly possesses all the required continuous and discrete symmetries, inclusive of parity (P) and time reversal (T),
\begin{equation}
L=-\frac{1}{2}\lambda^{-1}c\left(\frac{dx^\mu}{d\tau}\right)^2-\frac{1}{2}\lambda m^2 c - \frac{d x^\mu}{d\tau}\left(q\,A_\mu + g\, C_\mu\right)
+\frac{1}{2}\beta_3 \frac{d x^\mu}{d\tau}\,x^\nu\,\epsilon_{\mu\nu\lambda\eta}\left(g\,\partial^\lambda A^\eta - q\,\partial^\lambda C^\eta\right).
\end{equation}
This form also makes it explicit that $\beta_3$ is the sole physical parameter in SO(2) dual electrodynamics which couples the electric and magnetic
sectors of such systems. If $\beta_3=0$, these two electric and magnetic ``worlds'' live side by side without any interaction with one another whatsoever
through whatever interaction of an electromagnetic character---these two worlds are then totally ``dark'' to one another.
A nonvanishing $\beta_3$ parameter on the other hand, thus opens a portal between these two worlds,
basically through the interaction of a point electric charge in one world with a point magnetic monopole in the other world
(of course there may exist as well particles carrying both types of charges).

When considering an ensemble of $N$ such electrically and magnetically charged particles, {\sl a priori} one could envisage that for each particle the
parameter $\beta_3$ could take a different value, $\beta_{3,(s)}$, by which it would couple to the dual gauge fields $A^\mu$ and $C^\mu$. However
the latter fields in turn, find their origin in source currents $J^\mu_{e,m}$ themselves comprised of the contributions of other collections of such
electrically and magnetically charged particles. Hence finally all these particles will be in interaction through the exchange of the field strengths and their duals
of the gauge fields $A^\mu$ and $C^\mu$. From that more encompassing perspective, and because of the necessary
conservation of the total momentum of such dual electrodynamical systems (or equivalently Newton's Third Law in the nonrelativistic limit), the reciprocity
inherent to the concept of action-reaction implies that all values for the $\beta_{3,(s)}$ couplings coincide with a single parameter value, $\beta_3$.

Given this last remark, it is now possible to display the total classical action for such dual electrodynamics of systems of relativistic electrically and magnetically
charged massive point particles, which is manifestly invariant under SO(2) dual transformations as well as Poincar\'e and Lorentz transformations, which is
invariant under the local double gauge symmetries of dual electromagnetism, and which is invariant as well under parity and time reversal discrete symmetries.
And finally, which does not suffer from the presence of nonlocal singular structures for its gauge fields in the character of time-like Dirac strings
in vector gauge potentials. This action may be expressed as,
\begin{eqnarray}
&&S[A^\mu(x^\mu),C^\mu(x^\mu), x^\mu_{(s)}(\tau_s), \lambda_s(\tau_s)]= \nonumber \\
&=&\int d^4x^\mu\left\{-\frac{1}{4\mu_0 c}\left(\partial_\mu A_\nu - \partial_\nu A_\mu\right)^2
-\frac{1}{4\mu_0 c}\left(\partial_\mu C_\nu - \partial_\nu C_\mu\right)^2-\frac{1}{c}A_\mu J^\mu_e - \frac{1}{c} C_\mu J^\mu_m\right\} \nonumber \\
&& + \sum_s\int d\tau_s\left(-\frac{1}{2}\lambda^{-1}_s(\tau_s) c \left(\frac{dx^\mu_{(s)}(\tau_s)}{d\tau_s}\right)^2
- \frac{1}{2} \lambda_s(\tau_s) m^2_s c \right),
\end{eqnarray}
in which, through a procedure akin to the one outlined in the Appendix but now applied to the above final identification for the interacting Lagrangian $L_{int}$
(and by ignoring some total $\tau_s$-derivatives), the electric and magnetic current densities are identified to be given as,
\begin{eqnarray}
&&J^\mu_e(ct,\vec{x}\,)= \\
&=&\sum_s\left[ q_s c\frac{d x^\mu_{(s)}(t)}{d(ct)}\delta^{(3)}(\vec{x}-\vec{x}_{(s)}(t))-\frac{1}{2}\beta_3\, g_s c\  \partial_\nu\left(
\epsilon^{\mu\nu\rho\sigma}\frac{dx_{(s)\rho}(t)}{d(ct)}\,x_{(s)\sigma}(t) \delta^{(3)}(\vec{x}-\vec{x}_{(s)}(t))\right)\right], \nonumber
\end{eqnarray}
as well as,
\begin{eqnarray}
&&J^\mu_m(ct,\vec{x}\,)=  \\
&=&\sum_s\left[ g_s c\frac{d x^\mu_{(s)}(t)}{d(ct)}\delta^{(3)}(\vec{x}-\vec{x}_{(s)}(t))-\frac{1}{2}\beta_3\, q_s c\ \partial_\nu\left(
\epsilon^{\mu\nu\rho\sigma}\frac{dx_{(s)\rho}(t)}{d(ct)}\,x_{(s)\sigma}(t) \delta^{(3)}(\vec{x}-\vec{x}_{(s)}(t))\right)\right]. \nonumber
\end{eqnarray}
It may readily be verified that these currents are indeed conserved as it should, $\partial_\mu J^\mu_{e,m}=0$.

Note well that in the presence of a nonvanishing coupling constant $\beta_3$, the electric and magnetic charge and current densities
$J^\mu_{e,m}$, {\sl a priori} each associated to either the electric or the magnetic sector of the system separately, include then each a nonvanishing contribution
stemming from the point charges, whether $g_s$ or $q_s$, of the other sector. Through the interaction coupling in $\beta_3$, each of the two sectors is
no longer totally ``dark'' to the other sector; a nonvanishing coupling $\beta_3$ allows for some electromagnetic ``leakage'' or ``channeling'' of one sector
into the other.

To conclude, let us make explicit the 3-vector form of the Lorentz force equation (\ref{eq:Lorentz-final}) for the relativistic momentum,
$\vec{p}(t)=m\vec{v}(t)/\sqrt{1-\vec{v}\,^2(t)/c^2}$,
of a massive point particle of electric charge $q$ and magnetic charge $g$. A careful evaluation of the r.h.s.~of (\ref{eq:Lorentz-final})
leads to the expression,
\begin{eqnarray}
\frac{d}{dt}p^i(t) &=&  qc\left(\frac{\vec{E}_e}{c}+\frac{\vec{v}}{c}\times\vec{B}_e\right)^i\,+\,gc\left(\frac{\vec{E}_m}{c}+\frac{\vec{v}}{c}\times\vec{B}_m\right)^i
\nonumber \\
&+& \beta_3\,gc\left(\vec{B}_e-\frac{\vec{v}}{c}\times\frac{\vec{E}_e}{c}\right)^i\,-\,\beta_3\, qc\left(\vec{B}_m-\frac{\vec{v}}{c}\times\frac{\vec{E}_m}{c}\right)^i
\nonumber \\
&-&\frac{1}{2}\beta_3\,g\,\left[ - c x^j\frac{\partial \vec{B}^j_e}{\partial x^i} + ct\,\frac{d x^j}{dt}\,\frac{\partial \vec{B}^j_e}{\partial x^i} + \left(\vec{x}\times\frac{d\vec{x}}{dt}\right)^j\,
\frac{\partial \vec{E}^j_e/c}{\partial x^i} \right. \nonumber \\
&&\qquad\quad \left.
-ct\,\frac{d}{dt} \vec{B}^i_e(t,\vec{x}(t)) + \left(\vec{x}(t)\times\frac{d}{dt}\frac{\vec{E}_e(t,\vec{x}(t))}{c}\right)^i\ \right] \nonumber \\
&+&\frac{1}{2}\beta_3\,q\,\left[ - c x^j\frac{\partial \vec{B}^j_m}{\partial x^i} + ct\,\frac{d x^j}{dt}\,\frac{\partial \vec{B}^j_m}{\partial x^i} + \left(\vec{x}\times\frac{d\vec{x}}{dt}\right)^j\,
\frac{\partial \vec{E}^j_m/c}{\partial x^i} \right. \nonumber \\
&&\qquad\quad \left.
-ct\,\frac{d}{dt} \vec{B}^i_m(t,\vec{x}(t)) + \left(\vec{x}(t)\times\frac{d}{dt}\frac{\vec{E}_m(t,\vec{x}(t))}{c}\right)^i\ \right].
\label{eq:Lorentz-3vector}
\end{eqnarray}
Of course upon setting $g=0$ and $\vec{E}_m/c=\vec{0}=\vec{B}_m$, one recovers the ordinary Lorentz force equation of motion, even when $\beta_3\ne 0$.
Note as well, once again, how a nonvanishing value for the coefficient $\beta_3$ indeed allows for a interaction coupling between  the electric and
magnetic sectors of SO(2) dual electrodynamics.

\textcolor{black}{
On the other hand if $\beta_3=0$, the above generalised Lorentz force equation (\ref{eq:Lorentz-3vector}) reduces drastically, to the much simpler form,
\begin{equation}
\beta_3=0:\qquad
\frac{d}{dt}\vec{p} =  qc\left(\frac{\vec{E}_e}{c}+\frac{\vec{v}}{c}\times\vec{B}_e\right)\,+\,gc\left(\frac{\vec{E}_m}{c}+\frac{\vec{v}}{c}\times\vec{B}_m\right).
\label{eq:Lorentz-3vector-0}
\end{equation}
This expression certainly does not coincide with the choice usually made (arbitrarily) for a SO(2) dual covariant generalised Lorentz force, namely in the form
of (\ref{eq:generLorentz}) (without the indicated question mark), namely
\begin{equation}
\frac{d}{dt}\vec{p}=qc\left(\frac{\vec{E}}{c}+\frac{\vec{v}}{c}\times\vec{B}\right)\,+\,
gc\left(\vec{B}-\frac{\vec{v}}{c}\times\frac{\vec{E}}{c}\right).
\label{eq:generLorentz-0}
\end{equation}
Note that by using the relations (\ref{eq:E-B-em}) between $(\vec{E},\vec{B})$ and $(\vec{E}_{e,m},\vec{B}_{e,m})$, the latter form of generalised
Lorentz force is expressed as follows in terms of the doubled ensemble of electromagnetic fields in the present SO(2) dual invariant formulation,
\begin{eqnarray}
\frac{d}{dt}\vec{p} &=&
\ \ \ qc\left(\frac{\vec{E}_e}{c}+\frac{\vec{v}}{c}\times\vec{B}_e\right)\,+\,gc\left(\frac{\vec{E}_m}{c}+\frac{\vec{v}}{c}\times\vec{B}_m\right) \nonumber \\
&& +\ gc\left(\vec{B}_e - \frac{\vec{v}}{c}\times\frac{\vec{E}_e}{c}\right)\,-\,qc\left(\vec{B}_m - \frac{\vec{v}}{c}\times\frac{\vec{E}_m}{c}\right),
\end{eqnarray}
of which only the first line coincides with the generalised Lorentz force equation as obtained in (\ref{eq:Lorentz-3vector-0}) which in itself does derive
from a local Lorentz invariant variational principle, while the choice in (\ref{eq:generLorentz-0}) certainly does not \cite{Rohr}.}

\section{A Pair of Interacting Dyons in SO(2) Dual Electrodynamics}
\label{Sect4}

At the classical level the dimensionless coupling coefficient $\beta_3$ may {\sl a priori} take any real value. In order to explore the situation
within the quantum context and consider the possibility of a Dirac-Schwinger-Zwanziger quantisation condition \cite{Dirac1,Dirac2,Schwinger,Zwanziger}
for electric and magnetic point charges, if any,
let us now apply the above formulation of SO(2) dual invariant electrodynamics to a system composed of two distinct dyons, of charges $(q_0,g_0)$
and $(q,g)$. The dyon of charges $(q_0,g_0)$ is assumed to be static---thus implicitly of arbitrarily large mass---and positioned at the origin
of the inertial frame in use. The dyon of charges $(q,g)$, and of mass $m$, is left to propagate subjected to the generalised Lorentz force
identified above, exerted by the static and spherically symmetric electromagnetic fields sourced by the dyon of charges $(q_0,g_0)$.

Given the electric and magnetic 4-current densities associated to the dyon of charges $(q_0,g_0)$,
\begin{equation}
J^\mu_e(ct,\vec{x}\,)=(q_0 c,\vec{0}\,)\,\delta^{(3)}(\vec{x}\,),\qquad
J^\mu_m(ct,\vec{x}\,)=(g_0 c,\vec{0}\,)\,\delta^{(3)}(\vec{x}\,),
\end{equation}
in the Lorenz gauge with static gauge potentials then with components
$A^\mu(\vec{x}\,)=(A^0(\vec{x}\,),\vec{0}\,)$ and $C^\mu(\vec{x}\,)=(C^0(\vec{x}\,),\vec{0}\,)$,
the Maxwell equations for dual electromagnetism reduce to the following two Poisson equations,
\begin{equation}
-\vec{\nabla}^2 A^0(\vec{x}\,)=\mu_0 c\,q_0\,\delta^{(3)}(\vec{x}\,),\qquad
-\vec{\nabla}^2 C^0(\vec{x}\,)=\mu_0 c\,g_0\,\delta^{(3)}(\vec{x}\,),
\end{equation}
with the obvious solutions
\begin{equation}
A^0(\vec{x}\,)=\frac{\mu_0 c}{4\pi}\,q_0\,\frac{1}{r},\qquad
C^0(\vec{x}\,)=\frac{\mu_0 c}{4\pi}\,g_0\,\frac{1}{r},\qquad
r\equiv \sqrt{\vec{x}\,^2}.
\end{equation}
Correspondingly, one finds for the $\vec{E}_{e,m}$ and $\vec{B}_{e,m}$ fields in the electric and magnetic sectors,
\begin{equation}
\frac{\vec{E}_e(\vec{x}\,)}{c}=\frac{\mu_0 c}{4\pi}\,q_0\,\frac{\vec{x}}{r^3},\qquad
\vec{B}_e=\vec{0},\qquad
\frac{\vec{E}_m(\vec{x}\,)}{c}=\frac{\mu_0 c}{4\pi}\,g_0\,\frac{\vec{x}}{r^3},\qquad
\vec{B}_m=\vec{0},
\end{equation}
as well as for the final electric and magnetic fields, $\vec{E}$ and $\vec{B}$,
\begin{equation}
\frac{\vec{E}(\vec{x}\,)}{c}=\frac{\vec{E}_e(\vec{x}\,)}{c}=\frac{\mu_0 c}{4\pi}\,q_0\,\frac{\vec{x}}{r^3},\qquad
\vec{B}(\vec{x}\,)=\frac{\vec{E}_m(\vec{x}\,)}{c}=\frac{\mu_0 c}{4\pi}\,g_0\,\frac{\vec{x}}{r^3}.
\end{equation}

Upon substitution into the first two lines on the r.h.s.~of (\ref{eq:Lorentz-3vector}), one then finds for that contribution to the total generalised Lorentz force,
\begin{equation}
\frac{\mu_0 c^2}{4\pi}\,(q q_0 + g g_0)\,\frac{\vec{x}}{r^3}\,+\,\beta_3\,\frac{\mu_0 c}{4\pi}\,(qg_0 - g q_0)\,\frac{d\vec{x}}{dt}\times\frac{\vec{r}}{r^3},
\end{equation}
while the total contribution of the last four lines on the r.h.s.~of (\ref{eq:Lorentz-3vector}) reduces to,
\begin{equation}
\beta_3\,\frac{\mu_0 c}{4\pi}\,(q g_0 - g q_0)\,\frac{\vec{x}\times d\vec{x}/dt}{r^3}.
\end{equation}
The total Lorentz force equation of motion for the dyon of charges $(q,g)$ and mass $m$ thus reads,
\begin{equation}
\frac{d\vec{p}}{dt}=\frac{\mu_0 c^2}{4\pi}\,(q q_0 + g g_0)\,\frac{\vec{x}}{r^3},\qquad
\vec{p}=\frac{m\vec{v}}{\sqrt{1-\vec{v}\,^2/c^2}},\qquad
\vec{v}=\frac{d\vec{x}}{dt},
\end{equation}
namely a result which is totally independent of the coupling coefficient $\beta_3$ between the electric and magnetic sectors,
whatever the values for the charges $(q_0,g_0)$ and $(q,g)$.

On account of the time independency of the fields $\vec{E}_{e,m}$ and $\vec{B}_{e,m}$, on the one hand, and of their spatial spherical symmetry,
on the other hand, both the total energy and the total angular-momentum of the system are conserved observables,
\begin{equation}
\frac{d E}{dt}=0,\qquad
\frac{d\vec{L}}{dt}=\vec{0},
\end{equation}
with the values,
\begin{equation}
E=\frac{mc^2}{\sqrt{1-\vec{v}\,^2(t)/c^2}}\,+\,\frac{\mu_0 c^2}{4\pi}\,(q q_0 + g g_0)\,\frac{1}{r(t)},\qquad
\vec{L}=\vec{x}(t)\times\vec{p}(t).
\end{equation}
In particular note how the conserved angular-momentum is solely contributed by the orbital angular-momentum of the dyon of mass $m$.

Since the independency on the coupling constant $\beta_3$ of the above Lorentz force in this case may somewhat come as a surprising feature,
let us reconsider the situation now from a more general perspective. Again in the Lorenz gauge, imagine an arbitrary static configuration of the gauge 4-potentials
in the form of $A^\mu(\vec{x}\,)=(A^0(\vec{x}\,)=\Phi_A(\vec{x}\,)/c,\vec{0}\,)$ and $C^\mu(\vec{x}\,)=(C^0(\vec{x}\,)=\Phi_C(\vec{x}\,)/c,\vec{0}\,)$,
with their associated fields $\vec{E}_{e,m}(\vec{x}\,)$, while $\vec{B}_{e,m}=\vec{0}$,
\begin{equation}
\frac{\vec{E}_e(\vec{x}\,)}{c}=-\vec{\nabla} A^0(\vec{x}\,)=-\frac{1}{c}\vec{\nabla}\Phi_A(\vec{x}\,),\qquad
\frac{\vec{E}_m(\vec{x}\,)}{c}=-\vec{\nabla}C^0(\vec{x}\,)=-\frac{1}{c}\vec{\nabla}\Phi_C(\vec{x}\,).
\end{equation}
The Lagrangian for the dyon of mass $m$ in this choice of background fields then reduces to,
\begin{eqnarray}
&&L\left(x^\mu,\frac{dx^\mu}{d\tau};\lambda\right) = \nonumber \\
&=& -\frac{1}{2}\lambda^{-1}c\left(\frac{d x^\mu}{d\tau}\right)^2-\frac{1}{2}\lambda m^2 c
-\frac{d x^0}{d\tau}\left(q A^0 + g C^0\right)\,+\,\frac{1}{2}\beta_3\,\left(\vec{x}\times\frac{d\vec{x}}{d\tau}\right)\cdot
\left(q \frac{\vec{E}_m}{c} - g \frac{\vec{E}_e}{c}\right),\qquad
\end{eqnarray}
while using the proper-time world-line parametrisation that same action then reads,
\begin{equation}
S[\vec{x}\,]=\int dt\left[-mc^2\sqrt{1-\frac{1}{c^2}\left(\frac{d\vec{x}}{dt}\right)^2}\,-\,\left(q \Phi_A+g \Phi_C\right)
\,+\,\frac{1}{2}\beta_3\left(\vec{x}\times\frac{d\vec{x}}{dt}\right)\cdot\left(q\frac{\vec{E}_m}{c}-g\frac{\vec{E}_e}{c}\right)\right].
\end{equation}

Quite obviously now, whenever the fields $\vec{E}_e(\vec{x}\,)$ and $\vec{E}_m(\vec{x}\,)$ are spherically symmetric, as are those
of the dyon of charges $(q_0,g_0)$ with
\begin{equation}
\frac{\vec{E}(\vec{x}\,)}{c}=\frac{\vec{E}_e(\vec{x}\,)}{c}=\frac{\mu_0 c}{4\pi}\,q_0\,\frac{\vec{x}}{r^3},\qquad
\vec{B}(\vec{x}\,)=\frac{\vec{E}_m(\vec{x}\,)}{c}=\frac{\mu_0 c}{4\pi}\,g_0\,\frac{\vec{x}}{r^3},
\end{equation}
as well as $\Phi_A(\vec{x}\,)=\mu_0 c^2 q_0/(4\pi r)$ and $\Phi_C(\vec{x}\,)=\mu_0 c^2 g_0/(4\pi r)$,
this action becomes totally independent of the coupling $\beta_3$.

Not only does this general observation directly explain the result for the above Lorentz force in this case, but it also points to the following fact.
Keeping to the case of a pair of dyons, in order to obtain some dependency on the coupling $\beta_3$ for the trajectory of the dyon of mass $m$,
it is necessary that at last the fields $\vec{E}_{e,m}$ not be spherically symmetric (in some inertial frame), which requires that the source dyon
of charges $(q_0,g_0)$ be accelerated relative to any arbitrarily chosen inertial reference frame.
Consequently, effects arising from a nonvanishing coupling $\beta_3\ne 0$ are expected to be suppressed by factors in $\vec{v}_0/c$, $\vec{v}_0$ being
the velocity of the dyon of charges $(q_0,g_0)$. While in turn, back reaction effects of the fields generated by the dyon of mass $m$ on the trajectory of the
dyon of charges $(q_0,g_0)$ itself ought to be included as well, in order to assess the relative importance of effects generated by the coupling $\beta_3$
between the electric and magnetic sectors of this electrodynamical system, and prospects for the experimental determination of its value.

Alternatively one may consider a system made of three dyons, of which
the trajectory of one in the presence of the other two is closely studied, to determine possible effects generated by the coupling constant $\beta_3$
in combination with the presence of monopole magnetic charges.

Finally, it is worth contrasting the above situation for a pair of dyons with that which applies when rather the arbitrary choice of Lorentz force
in (\ref{eq:Lorentz-naive},\ref{eq:Lorentz-naive2}) is made, which cannot derive from a variational principle.
Given the above field configurations in $\vec{E}$ and $\vec{B}$ generated by the dyon
of charges $(q_0,g_0)$, the relativistic equation of motion for the second dyon of mass $m$ then reads,
\begin{equation}
\frac{d\vec{p}}{dt}=\frac{\mu_0 c^2}{4\pi}\,(q q_0 + g g_0)\,\frac{\vec{x}}{r^3}\,-\,\frac{\mu_0 c}{4\pi}\,(q g_0 - g q_0)\,\frac{\vec{x}\times d\vec{x}/dt}{r^3},
\end{equation}
leading again of course to a conserved total energy, $E$, and a conserved total angular-momentum, $\vec{M}$, now given by the expressions,
\begin{equation}
E=\frac{mc^2}{\sqrt{1-\vec{v}\,^2(t)/c^2}}\,+\,\frac{\mu_0 c^2}{4\pi}\,(q q_0 + g g_0)\,\frac{1}{r(t)},\quad
\vec{M}=\vec{x}(t)\times\vec{p}(t)-\frac{\mu_0 c}{4\pi}\,(q g_0 - g q_0)\,\frac{\vec{x}(t)}{r(t)}.
\end{equation}
In particular note well that for such a choice, besides the orbital angular-momentum $\vec{L}=\vec{x}\times\vec{p}$, the conserved total angular-momentum
$\vec{M}$ includes as well the electromagnetic contribution $-\mu_0c (q g_0 - g q_0)\hat{x}/4\pi$ (while the expression for the conserved relativistic total energy
is not modified from what it is above). Since within the context of a quantum mechanical description
the total angular-momentum $\vec{M}$ may be represented by quantum states spanning some SO(3) (rather more precisely, SU(2)) representation spaces
of integer and half-integer spin 
values $j$ (in units of $\hbar$), while the orbital angular-momentum may possess integer spin values only, necessarily this heuristic argument \cite{Jose,Alvarez}
(in the absence of a variational principle) requires the celebrated Dirac-Schwinger-Zwanziger quantisation condition \cite{Dirac1,Dirac2,Schwinger,Zwanziger}
to be met by the dyon charges $(q_0,g_0)$ and $(q,g)$, in the form of,
\begin{equation}
\frac{\mu_0 c}{4\pi}\,(q g_0 - g q_0) = \frac{1}{2}\hbar\,n,\qquad n\in\mathbb{Z}.
\end{equation}

Clearly such a conclusion does not apply to the formalism developed in this work, with in particular the generalised SO(2) dual invariant Lorentz force
uniquely identified herein from a variational principle required to reproduce the ordinary and experimentally confirmed Lorentz force in the absence
of monopole magnetic charges and currents. This approach implies the \textcolor{black}{possible} existence of a single dimensionless real coupling constant, $\beta_3$,
of presently unknown value, which channels interactions between the otherwise mutually (electromagnetically) totally ``dark'' electric and magnetic sectors
of SO(2) dual invariant electrodynamics.
Within this formalism monopole electric and magnetic charges need not meet a quantum Dirac-Schwinger-Zwanziger quantisation condition.
While there is no need either of time-like string singularities of the Dirac type for the doubled ensemble of 4-vector gauge potentials, $A^\mu(x^\mu)$ and $C(x^\mu)$,
in the presence of magnetic (or electric) monopoles.

\section{Conclusions}
\label{Sect5}

The present work has identified the space-time \textcolor{black}{strictly} local action principle for the SO(2) dual electrodynamics of systems of electrically and magnetically charged
point masses in interaction, based on the double requirements, on the one hand, that all space-time, SO(2) dual and gauge symmetries be manifest in
the Lagrangian action for such systems \textcolor{black}{and with fields devoid of any extended string-like
space-time singularities (except for those sourced by point charges)},
and on the other hand, that all the equations of ordinary Maxwell electrodynamics in the absence of magnetic monopole charges
which have been so soundly established by the verdict of so many experiments, are indeed perfectly reproduced when no magnetic charges nor currents
are present for SO(2) dual electrodynamics.

Except of course for the values of the electric and magnetic charges of \textcolor{black}{point} particles, this action for SO(2) dual electrodynamics
involves a single \textcolor{black}{possible} new interaction constant, which induces a coupling between the electric and magnetic sectors of any such system,
thereby opening a channel or portal between these two ``worlds'' which otherwise would remain totally ``dark'' to one another
through electromagnetic interactions. In addition, because of a doubled set of SO(2) dual 4-vector abelian gauge potentials, the presence of monopole
magnetic charges does not imply nonlocal Dirac string singularities for any of these gauge potentials. No Dirac-Schwinger-Zwanziger quantisation
condition applies either to electric and magnetic charges.

Furthermore, even though certainly SO(2) dual invariant, the generalised Lorentz force identified through the present analysis
is genuinely and physically truly different from the usual SO(2) dual invariant choice made heretofore arbitrarily just by simplicity, but which cannot derive
from some action principle. If one accepts the possible physical relevance of the present formulation of SO(2) dual electrodynamics,
this fact on its own requires an earnest reassessment of the conceptual methods, instrumental developments and experiments that have been
pursued over the decades in search of the existence of magnetic monopoles, as well as the identification of new strategies aiming towards the experimental
certification of this SO(2) dual formalism.

In particular, and somewhat as a surprise, it was observed that in order to unravel effects resulting from the coupling between the electric and magnetic sectors
of SO(2) dual electrodynamics, namely between electric and magnetic charges, and with the purpose as well of establishing the value of the extra coupling constant
between these two sectors, presumably a fully fledged analysis involving the acceleration and back-reaction of dyons in interaction is required.
Indeed in the case of a pair of dyons of which one is static, it turns out that the dynamics of the other is independent from that specific coupling constant.
The effects of that coupling constant would be suppressed by factors of dyon accelerations in units of $c$.

The present discussion also raises a variety of other questions and perspectives of interest and of physical relevance.
First one would like to extend SO(2) dual electrodynamics
to couple it to electrically and magnetically charged relativistic fields---rather than relativistic \textcolor{black}{point} particles---and explore the quantum properties
of these systems as relativistic quantum field theories, inclusive of the scattering, production and decay of their quantum particle excitations.

Another issue that obviously comes to mind is how to reconcile and embed this type of electrodynamics within the framework and concepts of
the Standard Model for quarks and leptons and (at least) their electro-weak interactions.

Clearly, the observed feature of mutually electromagnetically ``dark'' sectors of electrically and magnetically charged matter of course also raises
the issue of its possible relevance to the mystery of invisible or dark matter in the Universe. Could SO(2) dual electrodymamics provide
yet another candidate for dark matter, with a dark parallel world of purely magnetically charged matter that would interact electromagnetically
ever so weakly  with the purely electrically charged matter of our visible Universe, and yet contribute to the Universe's total energy-momentum tensor,
hence to its local curvature and thus eventually to the local dynamics of its visible matter?

Finally, in view of the gravito-electromagnetic approximation to General Relativity in linearised form, how could a parallel discussion be developed
for the gravitational interaction in its complete nonlinear realisation to account not only for gravito-electric charges, namely mass distributions
and their energy-momentum tensor as is achieved by General Relativity, but for gravito-magnetic charges as well, namely so-called NUT
(Newman-Unti-Tamburino) charges which are known to contribute as specific integration constants for certain classes of stationary axisymmetric
black hole solutions in General Relativity?

\textcolor{black}{
Besides these avenues to further investigate the physical relevance that the general approach outlined in the present work may possibly offer,
it also raises a series of issues in view of discussions available in the literature regarding magnetic monopoles, and in particular the Dirac monopole.
One first such point is that
of the lack of a Dirac-Schwinger-Zwanziger quantisation condition for dyonic charges. However in that respect, it should be pointed out that since
there is no action variational principle which leads to a generalised Lorentz force equation in the form of (\ref{eq:generLorentz-0}), the identification
of a conserved quantity which generalises the concept of an orbital angular-momentum in the presence of a point magnetic monopole whose
magnetic field is spherically symmetric, must rely on that generalised Lorentz force itself (rather than a Noether-like procedure based on an action),
so that such an eventual quantisation condition could indeed possibly be identified for the quantised system in the manner recalled above
even in the absence of an action principle.
In retrospect, it should thus not come as a surprise that by changing the generalised Lorentz force
equation as it would apply to dyons and their interactions, a likewise quantisation condition on dyons charges may or may not be implied.
With the generalised Lorentz force as identified in the present work but based indeed now on as general an action as possible given the constraints of
the experimentally confirmed ordinary electrodynamics, such a quantisation condition thus does not apply.}

\textcolor{black}{
The literature also presents specific classes of actions whose variation provide field equations
for the electromagnetic fields, including those contributed by Dirac magnetic monopoles with their space-time string-like singularities
(for reviews see Refs.\cite{Blago,Singleton,Milton}).
However in contradistinction to what is the case in the present work with no string-like singularities for fields and a strictly space-time local action
which at the classical level is manifestly Lorentz and SO(2) dual invariant, usually these discussions as developed in the literature \cite{Dirac2,Schwinger,Zwanziger,Milton,Zwanziger2} involve actions that classically are not manifestly
Lorentz invariant (nor strictly local), often because of the explicit contribution of some fixed constant background 4-vector breaking full Lorentz covariance and in direct
relation with the Dirac string characteristic of the Dirac monopole,
while at the quantum level Lorentz covariance may be restored provided however that dyonic charges meet a specific Dirac-Schwinger-Zwanziger
quantisation condition\cite{Milton,Zwanziger2}.
}

\textcolor{black}{
Examples of U(1)$\times$U(1) gauge theories are of course available in the literature\cite{Holdom}, or extensions of the Standard Model with an extra U(1)
field which mixes with the known photon and $Z_0$ fields\cite{Hook}. However these are not directly comparable to the type of U(1)$\times$U(1)
gauge theory involved in the present work, because of the extra global SO(2) duality symmetry imposed on the field content, inclusive of
its electric and magnetic charge content. In particular the matter field content to be considered within a quantum field theoretic context for what are the
point electric and magnetic charges of the present analysis, still needs to be properly identified, inclusive of its coupling to the gauge fields, and this
in correspondence with the generalised Lorentz force coupling that has been constructed herein in terms of point charges. It is only by completing
such a construction within a complete quantum field theoretic context that a proper physical interpretation of charges, fields and their interactions
can be achieved, in terms of electric and magnetic fields and charges. However, it is Maxwell's classical electrodynamics extended to SO(2) dual
electrodynamics that has taken us thus far already, in terms of the doubled set of gauge fields and their physical interpretation.}

\textcolor{black}{
Given the quantum field theoretic context and as already mentioned above, the extension of the present analysis to the Standard Model
with its SU(2)$\times$U(1) electroweak sector, and more generally to a spontaneously broken nonabelian gauge theory, also needs to be addressed
eventually. Within that latter context, when
a magnetic charge arises typically it then does enjoy a quantisation condition analogous to a Dirac-Schwinger-Zwanziger one.
However in that case, the {\sl raison d'\^etre} of this charge and its quantisation is purely topological\cite{Singleton},
with the source of the magnetic field then being spread over space because of an intertwined topologically nontrivial
configuration of scalar Higgs and vector gauge fields
each with their characteristic fall-off length scales.
The present discussion however, considers specifically point magnetic (and electric) charges, that should correspond to quanta of fields excitations,
rather than classical topologically nontrivial field configurations. In the case of such point charges, {\sl a priori} there is no necessity of some
charge quantisation condition of the Dirac-Schwinger-Zwanziger type, since no topological restriction is then involved.}

\textcolor{black}{
On par with other studies available in the vast literature on the ``magnetic monopole mystery'', the present alternative approach deserves as well
its own specific further investigations, in particular given its original new form of generalised Lorentz force equation to account for interacting dyons.}

\section*{Data Availability Statement}

This manuscript has no associated data.

\section*{Acknowledgement}

This work is supported in part by the {\sl Institut Interuniversitaire des Sciences Nucl\'eaires} (IISN, Belgium).

\section*{Appendix}
\label{Appendix}

Using the electromagnetic field strength components $F_{\mu\nu}$ as defined in (\ref{eq:Fmunu})
as well as $J^\mu_e=(c\rho_e,\vec{J}_e\,)$ which is such that $\partial_\mu\,J^\mu_e=0$, in manifest
Lorentz covariant form the ordinary Maxwell equations are compactly expressed as
\begin{equation}
\partial^\mu\,F_{\mu\nu}= \mu_0\,J_{e,\nu},\qquad
\partial^\mu\,{^*\!}F_{\mu\nu}=0,
\end{equation}
where ${^*\!}F_{\mu\nu}=\frac{1}{2}\epsilon_{\mu\nu\rho\sigma}\,F^{\rho\sigma}$.
By considering the 2-form $F=\frac{1}{2}dx^\mu\wedge dx^\nu\,F_{\mu\nu}$, its Hodge decomposition provides the following representation,
\begin{equation}
F_{\mu\nu}=\left(\partial_\mu A_\nu  - \partial_\nu A_\mu\right) \, - \, \frac{1}{2}\epsilon_{\mu\nu\rho\sigma}\left(\partial^\rho C^\sigma - \partial^\sigma C^\rho\right),
\end{equation}
where the vector potential 1-forms $A=dx^\mu\,A_\mu$ and $C=dx^\mu\,C_\mu$ are each defined up to an exact 1-form, namely up to some gauge transformation
in the form of
\begin{equation}
A'_\mu=A_\mu + \partial_\mu \Lambda_A,\qquad
C'_\mu=C_\mu + \partial_\mu \Lambda_C,
\end{equation}
$\Lambda_A(x^\mu)$ and $\Lambda_C(x^\mu)$ being two arbitrary scalar fields. Under the assumption that these fields vanish at infinity this gauge freedom
allows to require that the gauge potentials $A_\mu$ and $C_\mu$ are restricted to obey, for instance, the Lorenz gauge fixing condition,
\begin{equation}
\partial^\mu A_\mu=0,\qquad
\partial^\mu C_\mu=0
\end{equation}
(gauge transformations preserving these Lorenz gauge conditions must be such that $\square\Lambda_A=0$ and $\square\Lambda_C=0$).

In terms of these gauge potentials the above Maxwell equations are then of second-order in space-time derivatives,
\begin{equation}
\square A_\mu - \partial_\mu\left(\partial^\nu A_\nu\right)=\mu_0\,J_{e,\mu},\qquad
\square C_\mu - \partial_\mu\left(\partial^\nu C_\nu\right) = 0,\qquad
\square\equiv \partial_\mu\,\partial^\mu,
\end{equation}
which in the Lorenz gauge thus reduce to
\begin{equation}
\square A_\mu = \mu_0 J_{e,\mu},\qquad
\square C_\mu=0.
\end{equation}
Consequently when assuming that all fields $A_\mu$, $C_\mu$ and $F_{\mu\nu}$ vanish at infinity, necessarily
(because of the absence of a magnetic 4-current density $J^\mu_m$) the gauge potential $C_\mu$ vanishes identically (up to gauge transformations).
Hence one is left with the sole
gauge field $A_\mu$, leading to the usual representation\footnote{When the Maxwell equations are considered in explicit 3-form with the
electric and magnetic 3-vector fields $\vec{E}$ and $\vec{B}$, as is usually done, it is the Helmholtz decomposition for 3-vectors in 3-Euclidean space which readily
leads to the same conclusion in terms of a scalar potential $\Phi$ and a vector potential $\vec{A}$, with then $A^\mu=(\Phi/c,\vec{A}\,)$,
as well as $\vec{E}=-\vec{\nabla}\Phi-\partial_t\vec{A}$ and $\vec{B}=\vec{\nabla}\times\vec{A}$, as must follow then from the two ordinary
homogeneous Maxwell equations. Equivalently, the homogeneous equation $\partial^\mu\,{^*}F_{\mu\nu}=0$ means that the 2-form $F$ is closed, $d\wedge F=0$,
hence it can only be exact, such that $F=d\wedge A$, with thus $C^\mu=0$ up to gauge transformations.}
of the electric and magnetic fields, and of their field strength in the form of
\begin{equation}
F_{\mu\nu}=\partial_\mu A_\nu - \partial_\nu A_\mu.
\end{equation}

As is well known when expressed in terms of the vector potential $A_\mu$, the above second-order Maxwell equations derive from the following
space-time local and Poincar\'e invariant Lagrangian action principle in the field $A^\mu$ with source $J^\mu_e$,
\begin{equation}
S[A^\mu; J^\mu_e]=\int d^4x^\mu\,\left(-\frac{1}{4\mu_0 c}\left(\partial_\mu A_\nu - \partial_\nu A_\mu\right)^2\,-\,\frac{1}{c}A_\mu J^\mu_e\right),
\label{eq:SA}
\end{equation}
which is indeed gauge invariant (up to a surface term) on account of the conservation of the electric 4-current $J^\mu_e$.

Note that the associated manifestly Lorentz as well as gauge invariant first-order formulation is provided by the action in the following form,
\begin{equation}
S[A^\mu,\phi_{\mu\nu}; J^\mu_e]=\int d^4x^\mu\left(\frac{1}{4\mu_0 c}\phi^{\mu\nu} \phi_{\mu\nu}\,-\,
\frac{1}{2\mu_0 c}\phi_{\mu\nu}\left(\partial_
\mu A_\nu - \partial_\nu A_\mu\right) \, - \, \frac{1}{c}A_\mu J^\mu_e\right),
\end{equation}
where $\phi_{\mu\nu}$ is a real antisymmetric tensor field, whose equation of motion is $\phi_{\mu\nu}=\partial_\mu A_\nu - \partial_\nu A_\mu$.
Incidentally this is as well \cite{Govaerts1} precisely the Hamiltonian first-order action of the system, from which its phase space coordinates, Poisson brackets,
first-class Hamiltonian and first-class constraint (namely Gauss' Law) with its Lagrange multiplier may readily be identified \cite{Govaerts2}.

Let us now consider the situation when the electric 4-current $J^\mu_e$ is provided by a collection of point charges of masses $m_s$ and electric charges $q_s$,
each tracing its space-time trajectory $x^\mu_{(s)}(\tau_s)$ parametrised by a world-line coordinate $\tau_s$, such that $x^0_{(s)}(\tau_s(t))=ct$ in the specifically considered
inertial reference frame in which its 3-vector trajectory is observed to be $\vec{x}_{(s)}(t)=\vec{x}_{(s)}(\tau_s(t))$.

As is well known, and as may readily be checked, the free kinematics of such a collection of massive relativistic point particles is accounted for by the following sum
of their individual world-line actions,
\begin{equation}
S[x^\mu_{(s)},\lambda_s]=\sum_s c\int d\tau_s\left(-\frac{1}{2}\lambda^{-1}_s(\tau_s)  \left(\frac{dx^\mu_{(s)}(\tau_s)}{d\tau_s}\right)^2
- \frac{1}{2} \lambda_s(\tau_s) m^2_s \right),
\end{equation}
where $\lambda_s(\tau_s)>0$ is a (nowhere vanishing) world-line ein-bein for the particle of mass $m_s$, so that each of these actions is
invariant under (orientation preserving) diffeomorphisms in $\tau_s$ for its world-line\footnote{As a matter of fact, $\lambda_s(\tau_s)$ is as well
a Lagrange multiplier for the generator (and first-class constraint) of such world-line diffeomorphisms in the world-line coordinate $s$ \cite{Govaerts2}.}.
In particular a proper-time parametrisation corresponds to the choice
$\lambda_s(\tau_s)=(\dot{x}_{(s)}^2(\tau_s))^{1/2}/m_s$ (with $\dot{x}^\mu_{(s)}(\tau_s) \equiv dx^\mu_{(s)}(\tau_s)/d\tau_s$), as is implied by the variation
in $\lambda_s(\tau_s)$ of the above action.

The coupling of this system to the electromagnetic fields is effected through the term in $A_\mu J^\mu_e/c$ in (\ref{eq:SA}). For the electric 4-current
$J^\mu_e=(c\rho_e,\vec{J}_e)$, here let us make the usual choice with, 
\begin{equation}
\rho_e(ct,\vec{x})=\sum_s q_s\,\delta^{(3)}(\vec{x}-\vec{x}_{(s)}(t)),\qquad
\vec{J}_e(ct,\vec{x})=\sum_s q_s\frac{d\vec{x}_{(s)}(t)}{dt}\,\delta^{(3)}(\vec{x}-\vec{x}_{(s)}(t)),
\end{equation}
namely,
\begin{equation}
J^\mu_e(ct,\vec{x})=\sum_s q_s\frac{dx^\mu_{(s)}(\tau_s(t))}{dt}\delta^{(3)}(\vec{x}-\vec{x}_{(s)}(t)),
\end{equation}
which is indeed such that $\partial_t\rho_e+\vec{\nabla}\cdot\vec{J}_e=0$, or $\partial_\mu J^\mu_e=0$.
Given the identification of the gauge 4-potential $A^\mu=(\Phi/c,\vec{A})$ with the scalar and vector potentials $\Phi$ and $\vec{A}$, respectively,
one then has,
\begin{equation}
\int d^4 x^\mu\,\frac{-1}{c}J^\mu_e A_\mu=\sum_s \int dt\left(-\,q_s\Phi(ct,\vec{x}_{(s)}(t))\,+\,q_s\frac{d\vec{x}_{(s)}(t)}{dt}\cdot\vec{A}(ct,\vec{x}_{(s)}(t))\right).
\end{equation}
Given the changes of variables $ct(\tau_s)=x^0_{(s)}(\tau_s)$ (and conversely $x^0_{(s)}(\tau_s(t))=ct$), this same expression may be given in the form,
\begin{equation}
\int d^4 x^\mu\,\frac{-1}{c}J^\mu_e A_\mu=-\sum_s q_s\int d\tau_s\,\frac{dx^\mu_{(s)}(\tau_s)}{d\tau_s} A_\mu(x^\mu_{(s)}(\tau_s)).
\end{equation}

In conclusion the total action for such an electrodynamics system is given as,
\begin{eqnarray}
&&S[A_\mu;x^\mu_{(s)},\lambda_s] = \int d^4x^\mu\,\left(-\frac{1}{4\mu_0 c}\left(\partial_\mu A_\nu - \partial_\nu A_\mu\right)^2 \right) \nonumber \\
&& + \sum_s\int d\tau_s\left(-\frac{1}{2}\lambda^{-1}_s(\tau_s) c \left(\frac{dx^\mu_{(s)}(\tau_s)}{d\tau_s}\right)^2
- \frac{1}{2} \lambda_s(\tau_s) m^2_s c - q_s \frac{dx^\mu_{(s)}(\tau_s)}{d\tau_s} A_\mu(x^\mu_{(s)}(\tau_s))\right)\!.\ \ 
\end{eqnarray}
Besides the Maxwell equations in the form of
\begin{equation}
\partial_\mu F^{\mu\nu}(ct,\vec{x}) = \mu_0 J^\nu_e(ct,\vec{x})=\sum_s q_s\frac{dx^\nu_{(s)}(\tau_s(t))}{dt}\delta^{(3)}(\vec{x}-\vec{x}_{(s)}(t)),
\end{equation}
with $\partial_\mu\,{^*\!}F^{\mu\nu}=0$ since $F_{\mu\nu}=\partial_\mu A_\nu - \partial_\nu A_\mu$, it readily follows that one has the
following Lorentz force equations of motion,
\begin{equation}
\frac{d}{d\tau_s}\left(\frac{m_sc}{\sqrt{\dot{x}_{(s)}^2(\tau_s)}}\,\dot{x}_{(s),\mu}(\tau_s)\right)=q_s\,F_{\mu\nu}(x^\mu_{(s)}(\tau_s))\,\dot{x}^\nu_{(s)}(\tau_s),\qquad
\dot{x}^\mu_{(s)}(\tau_s)\equiv\frac{dx^\mu_{(s)}(\tau_s)}{d\tau_s}
\end{equation}
(the variations of the Lagrange multipliers $\lambda_s(\tau_s)$ imply the proper-time world-line parametrisation
solutions $\lambda_s(\tau_s)=((dx^\mu_{(s)}(\tau_s)/d\tau_s)^2)^{1/2}/m_s$),
or in explicit 3-vector form,
\begin{equation}
\frac{d}{dt}\vec{p}_{(s)}(t)=q_s\vec{E}(t,\vec{x}_{(s)}(t))+ q_s \vec{v}_{(s)}(t)\times\vec{B}(t,\vec{x}_{(s)}(t)),\qquad
\vec{p}_{(s)}(t)\equiv\frac{m_s\vec{v}_{(s)}(t)}{\sqrt{1-\vec{v}_{(s)}(t)/c^2}},
\end{equation}
with of course $\vec{v}_{(s)}(t)=d\vec{x}_{(s)}(t)/dt$.

\end{document}